\documentclass[a4paper,11pt]{article}
\pdfoutput=1
\usepackage{jheppub}
\usepackage{amsmath}
\usepackage{amsfonts}
\usepackage{amssymb}
\usepackage{graphicx}
\usepackage{wrapfig}
\usepackage{braket}
\usepackage{hyperref}
\usepackage{amsthm,verbatim,cancel}
\usepackage{color}
\usepackage{caption}
\usepackage{subcaption}
\newcommand{\pd}{\partial}

\newcommand{\ba}{\begin{align}}
\newcommand{\ea}{\end{align}}
\newcommand{\bea}{\begin{eqnarray}}
\newcommand{\eea}{\end{eqnarray}}
\newcommand{\nn}{\nonumber\\}

\DeclareMathOperator{\tr}{tr}
\renewcommand{\Re}{\operatorname{Re}}

\newcommand{\gabriel}{\textcolor{blue}}

\title{Gluing together Modular flows with free fermions}

\author{Gabriel Wong}
\emailAdd{gabrielwon@gmail.com}
\affiliation{Department of Physics, Fudan University, 2005 Songhu Road, 200438 Shanghai, China}

\abstract{We revisit the calculation of multi-interval modular Hamiltonians for free fermions using a Euclidean path integral approach.  We show how the multi-interval modular flow is obtained by gluing together the single interval modular flows.  Using this relation, we obtain an exact expression for the multi-interval modular Hamiltonian and entanglement entropy in agreement with existing results.   An essential ingredient in our derivation is the introduction of the \emph{modular action}. This determines the non-local field theory describing the free fermion reduced density matrix, and makes manifest it's non-local conformal symmetry and $U(1)$ Kacs-Moody symmetry. }
\begin{document}

\maketitle
\section{Introduction} 

A fundamental distinction between quantum and classical physics lies in the way information is lost upon restriction to a subsystem. Due to entanglement between subsystems the reduced density matrix can be mixed even when the global state is pure: thus perfect knowledge of the total system does not lead to perfect knowledge of it's subsystems.    The reduced density matrix completely characterizes the entanglement of a subsystem with it's environment, and has been a subject of intense study in both high energy and condensed matter physics.  In particular, the modular Hamiltonian- defined as the negative logarithm of the reduced density matrix- provides an effective Hamiltonian for the subsystem that plays a key role in the reconstruction of bulk operators in AdS/CFT \cite{Faulkner:2017vdd,Kabat:2017rw}, in the formulation of entanglement edge modes which distinguish different topological phases of matter \cite{Wong:2017pdm}, and in the proof of quantum null energy conditions \cite{Balakrishnan:2017bjg}. 

Given a spatial region $V$ and a state $\rho$ the reduced density matrix $\rho_{V}$ can be defined via a partial trace:
\bea
\rho_{V} =\tr_{\bar{V}} \rho
\eea
over the complement region $\bar{V}$, or by demanding that $\rho_{V}$ reproduces all expectation values of operators $O_{V}$ supported in $V$:
\bea \label{alg}
\braket{ O_{V}} = \tr_{V}(\rho_{V} O_{V} ) 
\eea
The Modular Hamiltonian $H_{V}$ is defined by 
\bea
\rho_{V}= \frac{e^{-H_{V}} }{Z_{V}}
\eea
where $Z_{V}= \tr e^{-H_{V}} $ is a normalization factor that can be interpreted as an effective partition function for the region $V$.
The modular flow is the evolution of operators $O_{V}$ inside the causal development of $V$ under $H_{V}$:
\bea
O_{V}(s) = e^{i H_{V} s} O_{V} e^{- i H_{V} s}
\eea 
Taking the logarithm of $\rho_{V}$ is a non-trivial operation so computing $H_{V}$ is difficult in the absence of symmetry constraints.  However, for the vacuum state reduced to a half space in a Lorentz invariant quantum field theory, $H_{V}$ can be identified with the boost generator \cite{bisognano1975duality}:  
Thus the modular flow defined by the evolution under $e^{- i t H_{V}}$ is \emph{geometric}: it evolves operators along accelerating trajectories inside the Rindler wedge.   In a conformal field theory, this result can be generalized to spherical regions, where the modular flow consists of geometric evolution inside a causal diamond \cite{myers}.  In either case, the modular Hamiltonian is a local operator related to the stress tensor $T_{ab}$, and can be expressed in the form
\bea \label{local} 
H_{V} = \int_{V} \, \beta(x) \, T_{00}  (x)
\eea
for some function $\beta (x)$, which acts as an inverse, \emph{local} temperature.   Some generalizations of the formula \eqref{local} can be obtained in quench states, in systems with boundary, and for regions lying on null planes \cite{Casini:2017roe}.  In all these cases the modular hamiltonian is directly related to the stress tensor.  

The most intuitive way to understand \eqref{local} is to represent the matrix elements $\braket{\phi_{f}|\rho_{V}|\phi_{i}}$ as a Euclidean path integral on a spacetime that is cut open along $V$, where  boundary conditions $\phi_{f}$ and $\phi_{i}$  are imposed.  Interpreting  $\braket{\phi_{f}|\rho_{V}|\phi_{i}}$ as a transition amplitude then provides a \emph{geometric} way to take the logarithm of $\rho_{V}$, provided there is a rotational symmetry around the entangling surface.  In this case, the rotational generator $K$ is conserved, and the Modular Hamiltonian can be identified with $2 \pi K$.   In the absence of such a rotational symmetry, we can still express  $\rho_{V}$ as a path-order exponential of $K$, but taking the logarithm is no longer an easy task.   

\begin{figure}
\centering
\includegraphics[scale=.2]{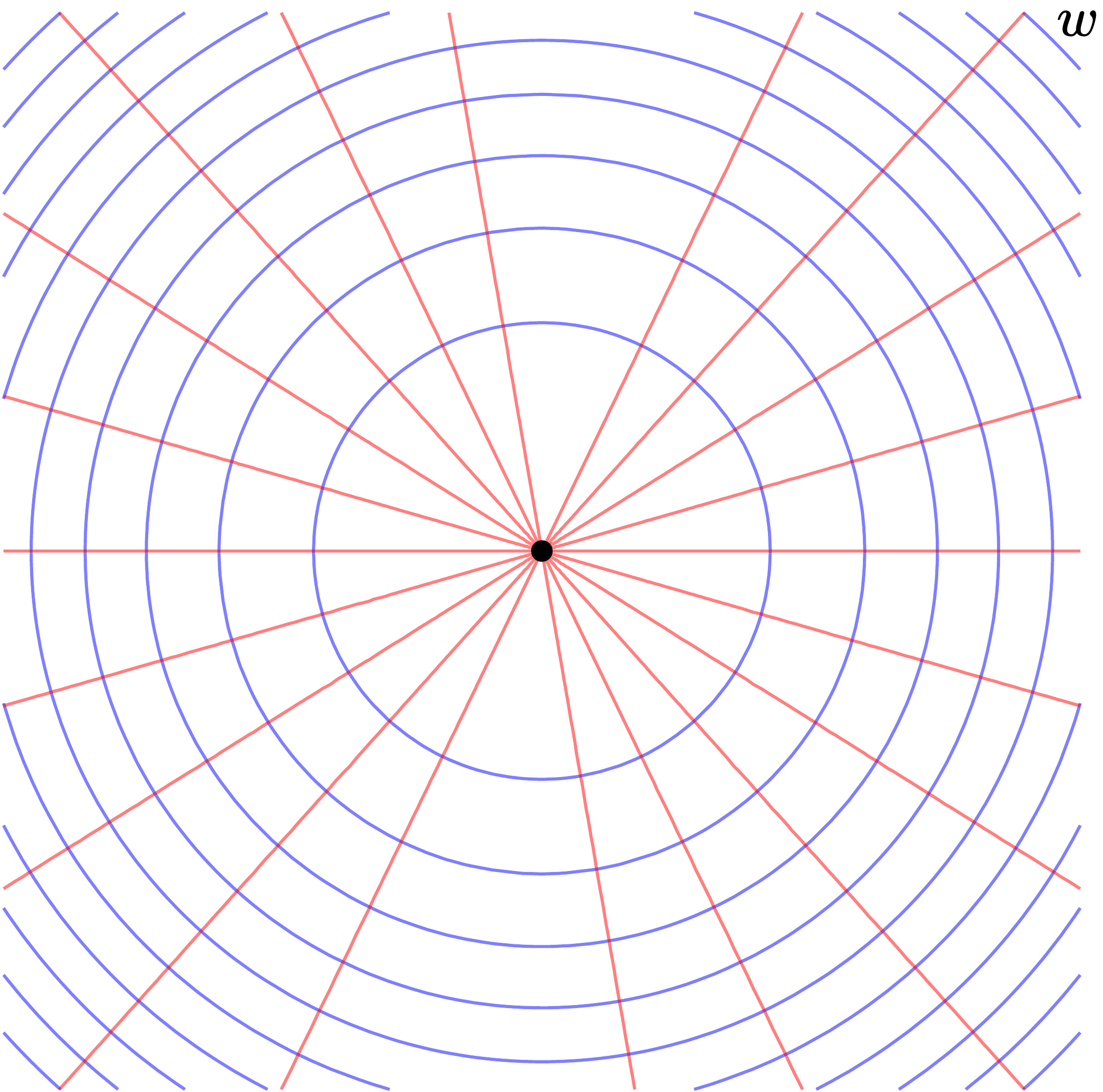}
\caption{The figure shows a foliation of the plane describing the half-line modular flow.   This corresponds to polar coordinates on the plane, with the angular direction determining the temperature circle }
\label{rot}
\end{figure}

For conformal field theories, there exist a (conformal) rotation symmetry around a single spherical or planar entangling surface $\pd V$ and \eqref{local} gives $H_{V}=2\pi K$ , with $K$ the generator of (conformal) rotations.  In particular, for  a 1+1 CFT on the complex $w$ plane, the modular hamiltonian for a half line  \eqref{local} takes the form
\bea \label{mod}
H_{\mathbb{R}^+}= \int_{V} 2 \pi  x T_{00} dx  =   \left(\int_{\mathbb{R}^+} 2 \pi  w T(w) dw +  \int_{\mathbb{R}^+} 2 \pi  w \bar{T}(\bar{w}) d\bar{w} \right)
\eea
The induced Euclidean modular flow corresponds to rotations around the origin and the constant modular time slices are rays leaving the origin (figure \ref{rot}) .   
Unfortunately when $V$ is no longer spherical or planar, or consists of multiple regions, the rotation symmetry around $\pd V$ is broken so \eqref{local} no longer holds.   It is therefore remarkable that for free chiral fermions in 1+1 dimension, Casini and Huerta \cite{casini2009reduced} were able to obtain an exact formula for the modular Hamiltonian when $V = \cup_{i} (a_{i},b_{i})$ consists of disjoint intervals on the real line and $\rho$ is the vacuum.   They used an algebraic approach based on solving \eqref{alg} directly for $\rho_{V}$ in terms of the two point function projected on $V$, and obtained a modular hamiltonian $H_{V}$ that induces an almost geometric flow : 
\begin{align}\label{CH}
 H_{V} &=2 \pi  i \int_{V}    \Psi^{\dagger}(z)  \frac{\delta (u(z)-u(y) 
)}{z-y}\Psi(y) \, dz \, dy \\
u(z)&= \log \prod_{i=1}^{n} \frac{z-a_{i}}{z-b_{i}}
\end{align}
Applying the standard delta function identity $ \delta (u(z)-u(y))= \sum_{\text{roots} \,y_{i}}  (\frac{du}{dy})^{-1}  \delta (z-y_{i})$, one sees a local term for roots $y_{i}(z)$ of $u$ living in the same interval as $z$, and a bilocal term for roots $y_{i}(z)$ living in other intervals.  
This bilocal term introduces mixing of the fermion fields on different intervals, on top of the geometric flow induced by the local term which is of the form \eqref{local}.  
This result was subsequently generalized to fermions on a circle with general boundary conditions in \cite{Klich:2015ina}, where it was shown that the computation of the modular Hamiltonian is related to the solution of a Riemann Hilbert problem.  

In this work we revisit the calculation of \eqref{CH} from the point of view of the Euclidean path integral.  We will find that the multi-interval modular flow can be obtained by gluing together single interval modular flows.  In the operator algebra language, this gluing is implemented by a non-local isomorphism between the single interval algebra of $n$ free fermions, and the $n$-interval algebra of a single fermion, as was first observed in \cite{rehren2013multilocal}. Our work provides a path integral interpretation of this isomorphism.   An important conceptual tool in our derivation is the modular action $S_{V}$, which determines a non-local finite temperature field theory for observers in the region $V$.  What makes the free fermion special is that this modular action also defines a CFT, except that the conformal symmetry is non-local and has a different central charge than the original theory.   We will provide a geometric derivation of the modular action,  from which the modular Hamiltonian \eqref{CH} will follow.  In section two, we illustrate how this works by first studying the two-interval case on the real line and then generalizing to multiple intervals and finite size.  In section three, we will show how the modular action makes manifest the non local symmetries of the multi-interval reduced density matrix.  This will provide a geometric interpretation of the non-local U(1) symmetry and conformal symmetry of the multi-interval operator algebra that was first observed in \cite{rehren2013multilocal}.  In section three, we make use of the invariance of the modular action under non-local conformal mappings to compute the entanglement entropy, and explain why it can be obtained by integrating a local entropy density.   This local entropy density in turn defines a local entanglement temperature, just as in the single interval case.   We will  conclude by discussing possible applications of these ideas to generic CFT's.
\section{ Multi-interval modular Hamiltonian for free Dirac Fermions} 
In this section, we will provide a path-integral derivation of the multi-interval modular Hamiltonian for free Dirac Fermions $\Psi, \bar{\Psi} $ living on the real line.   The Euclidean action on the complex $z$ plane is
\bea \label{global}
S =  \int \Psi^{\dagger}\bar{\pd}  \Psi  + \bar{\Psi}^{\dagger} \pd  \bar{\Psi}  \,\,d^{2}z .
\eea 
Henceforth, the treatment of the anti-holomorphic sector will be completely parallel to the holomorphic sector, so for brevity, we will occasionally omit the anti-holomorphic sector in the formulas below.   The action  \eqref{global} is the one appropriate to the local theory quantized on the real line.   However we will be interested the generically non-local action $S_{V}$ that defines a finite temperature path integral for $Z_{V}=\tr(e^{-H_{V}})$ .  This is a non-local field theory that determines the reduced density matrix in the sense that the thermal averages in $Z_{V}$ produces the expectation values with respect to $\rho_{V}$.  When $V$ is a the positive real axis on the complex plane, the temperature circle is the angular direction around the origin, and the modular action is the local Rindler action for uniformly accelerated Rindler observers.  In the case of disjoint intervals, the angular direction around the entangling surface encounters an apparent obstruction that invalidates a naive finite temperature interpretation.  In the following section we will show how to treat this apparent obstruction and obtain the \emph{modular} action $S_{V}$, from which we will derive the modular Hamiltonian \eqref{CH}.
\subsection{The two interval case} 
\begin{figure}
\centering
\includegraphics[scale=.4]{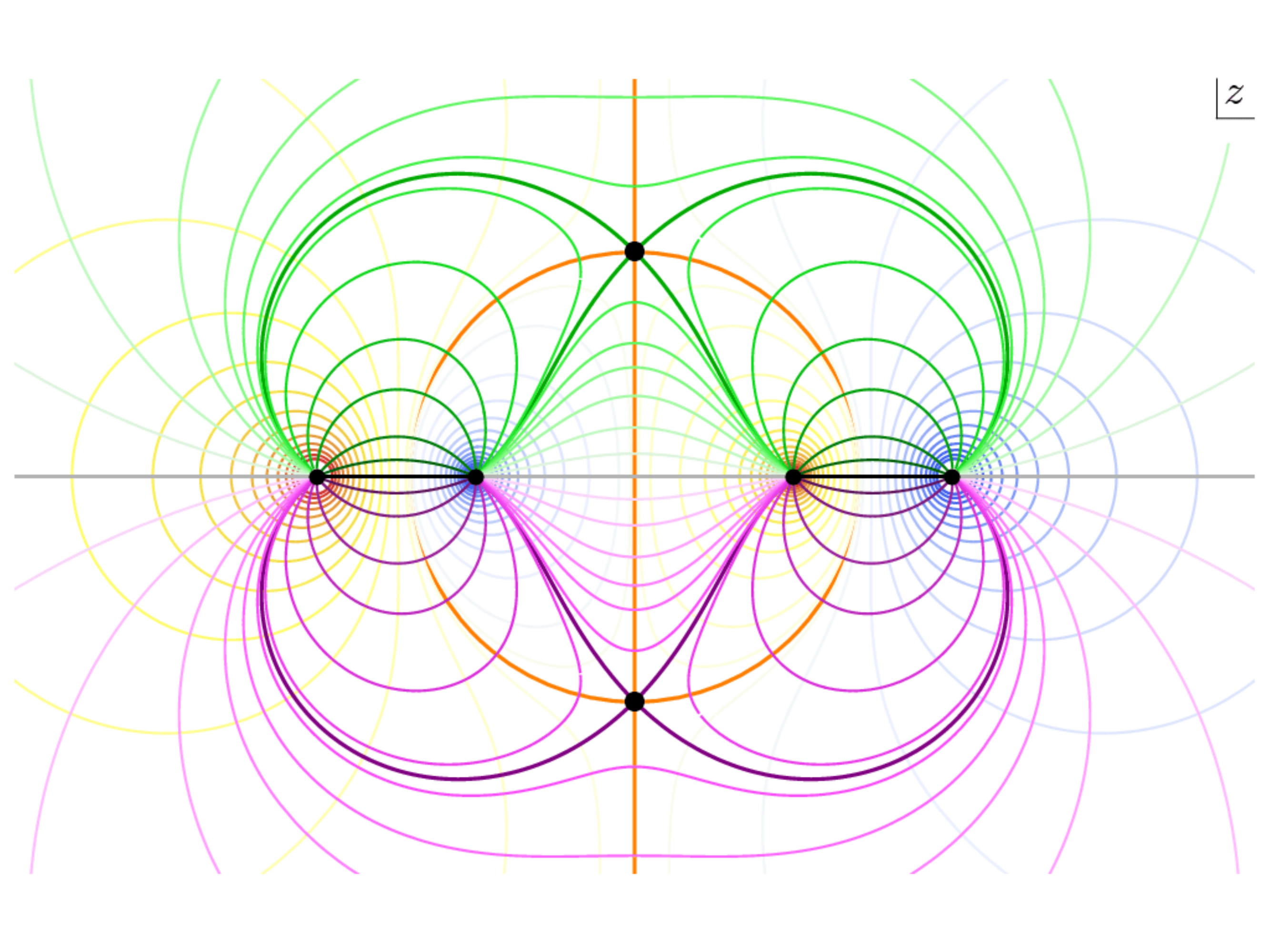}
\caption{Modular flow lines for 2 intervals, borrowed from \cite{Cardy:2016yq}.  The orange circle divides each of the intervals in the two halves,  which are cut re-glued when they evolve pass the singularities at  $z_{\pm}$ }
\label{flow}
\end{figure}

To keep things simple, we begin with two symmetrically placed intervals  $V = ( -b,-a) \cup (a,b) $ on the real axis of the complex $z$ plane.   The aforementioned obstruction can be nicely illustrated by trying to map the half-line modular flow of figure \ref{rot} into the two-interval modular flow of figure \ref{flow} via the mapping
\bea \label{w} 
w= f(z) = \frac{ (z+b)(z-a)}{(z+a)(z-b)},
\eea
as discussed in \cite{Cardy:2016yq}.  
This mapping is two to one, with branch point singularities at 
\bea
z_{\pm}= \pm i \sqrt{ab} 
\eea
where it is one to one and fails to be conformal because the derivative vanishes there.  This means that if we try to lift the half line modular flow on the $w$ plane to the $z$ plane we will find two singularities at $z_{\pm}$, in addition to the usual ones at the entangling surface $\pd V$.  The field lines of this two-interval modular flow is shown in figure \ref{flow}, borrowed from \cite{Cardy:2016yq}.  While the singularities at $\pd V$ can be regulated by cutting out small disks, the singularities at $z_{\pm}$ present an obstruction to the naive, local modular flow generated by \eqref{local}.  Indeed, if we carefully follow the evolution of the constant time slices (green and purple lines) in figure \ref{flow},  we see that as the two intervals approach the singularities $z_{\pm}$, they are cut and reconnected so that right endpoint of the left interval becomes connected with the left endpoint of the right interval and vice versa.  This led the the authors of \cite{Cardy:2016yq} to characterized the two- interval reduced density matrix $\rho_{V}$ by the sequence of operators: 
\bea \label{rhoV}
\rho_{V}= e^{-\theta_{0} K_{V} }  X^{\dagger} e^{ - (2 \pi -2 \theta_{0})K_{\bar{V}}} X e^{ -\theta_{0} K_{V}} 
\eea
Here $K_{V}$ and $ K_{\bar{V}}$ generate a geometric flow that evolve regions $V$ and $\bar{V}$ in the angular coordinate $\theta$ which runs orthogonal to $V$ and $\bar{V}$: these operators are given by the integral of the stress tensor over regions $V$ and $\bar{V}$ respectively, similar to \eqref{local}.  The ``cutting and regluing" operators $X$ and $X^{\dagger}$ are inserted at $z_{\pm}$: they map from the Hilbert space $\mathcal{H}_{V}$ to the complement $ \mathcal{H}_{\bar{V}}$ and vice versa, allowing the continuation of the modular flow past the singularities at $z_{\pm}$ by switching back and forth between evolution by $K_{V}$ and $ K_{\bar{V}}$.   To obtain the modular Lagrangian, we must incorporate the effects of $X$ and $X^{\dagger}$ into the path integral. Below we will show how this can be done by orbifolding the complex $z$ plane and imposing twisted boundary conditions.   These boundary conditions will generate the bilocal term in the Modular Hamiltonian \eqref{CH},  and can be interpreted as a gluing of the single interval modular flows that entangles the disjoint intervals.

\begin{figure}
\centering 
\includegraphics[scale=.20]{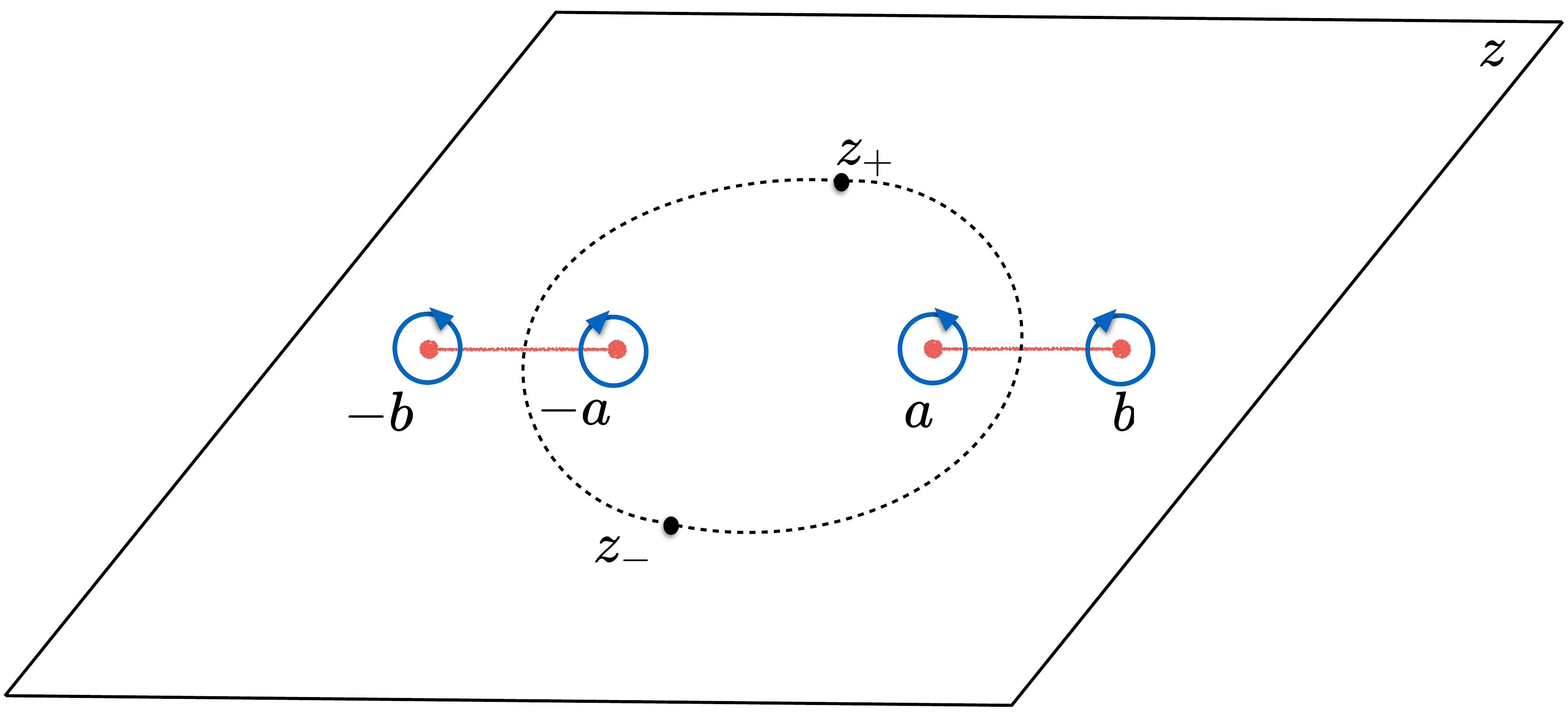}
\qquad \qquad
\includegraphics[scale=.25]{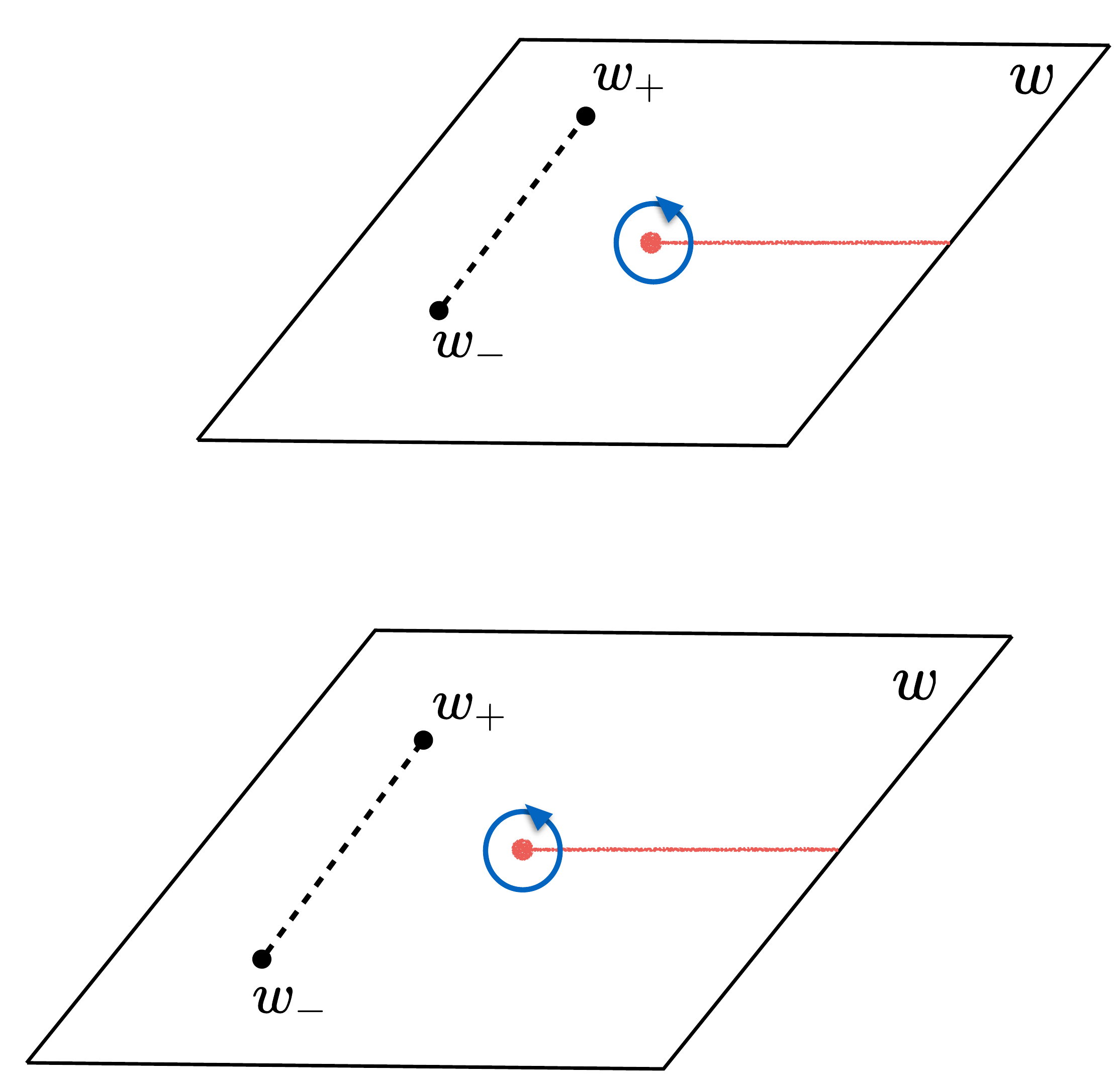}
\qquad \qquad
\caption{Shown are two coordinate systems describing the same reduced density matrix for two intervals.  On the left is the $z$ plane that has been cut open along the two intervals.  The blue circles are boundaries of disks that have been removed from the UV regulated spacetime, with the arrows indicating direction of the modular flow.  In the $w$ coordinates, the same background is represented as a double cover with branch points at $w_{\pm}$.}
\label{figure:surface}
\end{figure}

\begin{figure}
\centering 
\includegraphics[scale=.16]{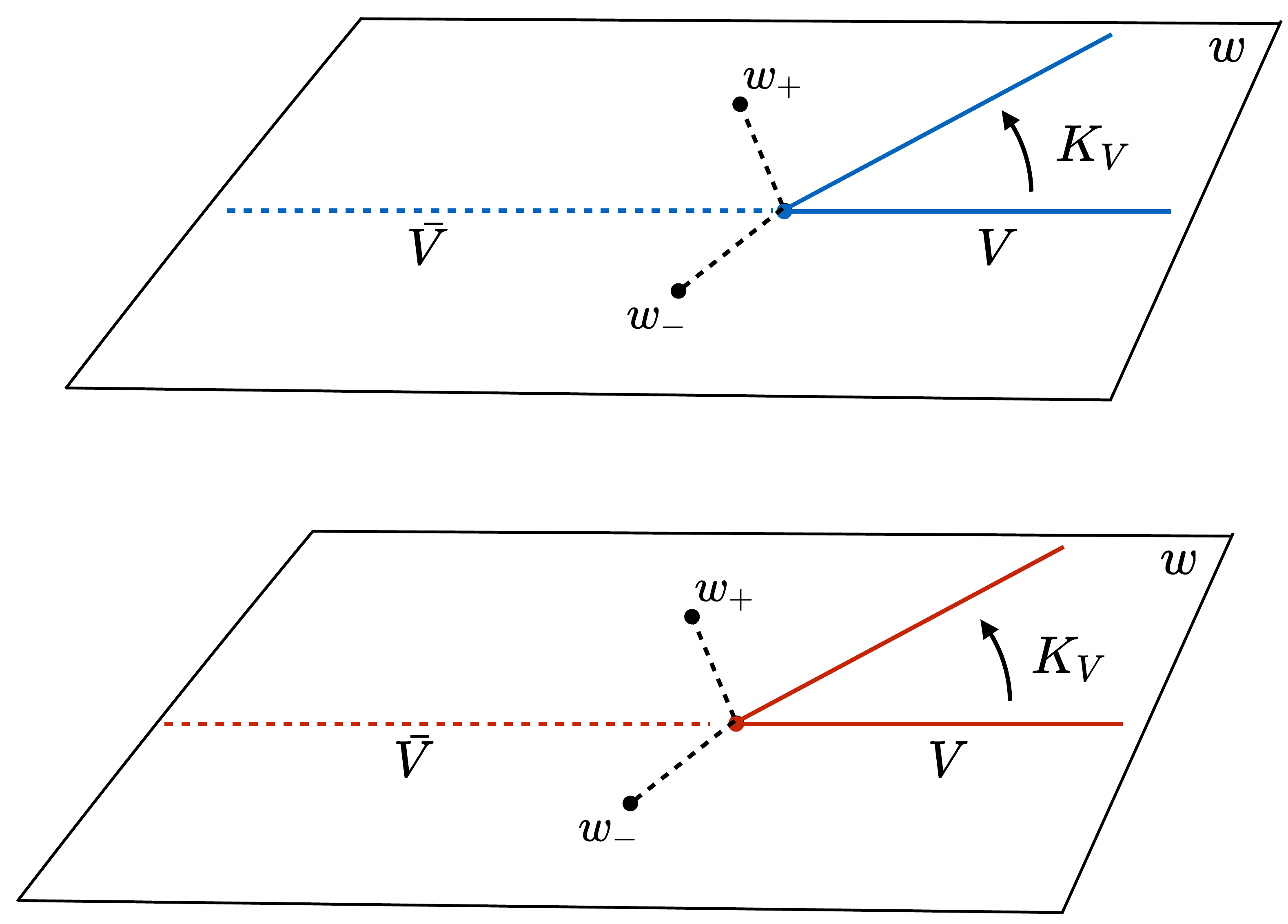}
\qquad 
\includegraphics[scale=.16]{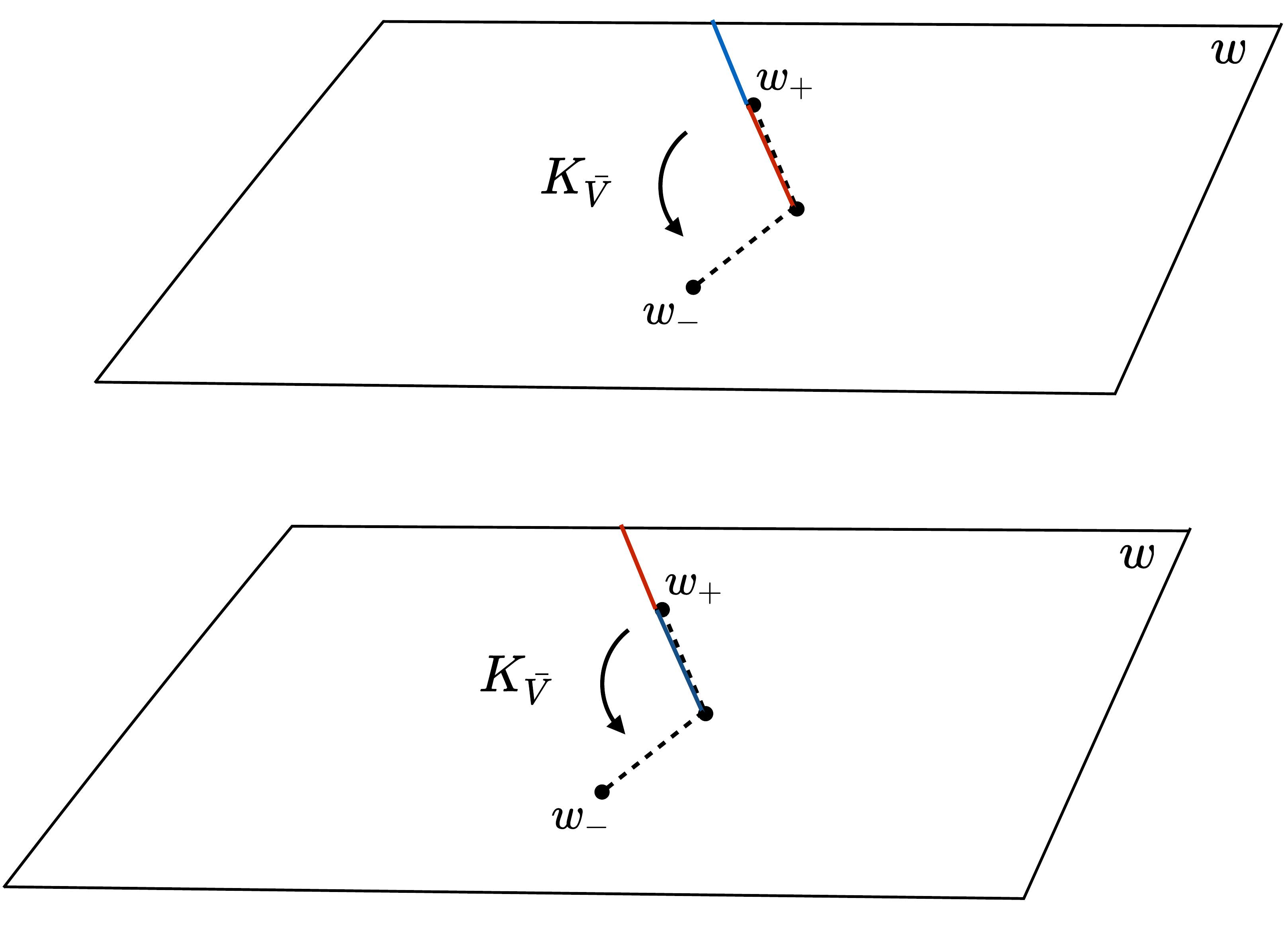}
\qquad 
\includegraphics[scale=.16]{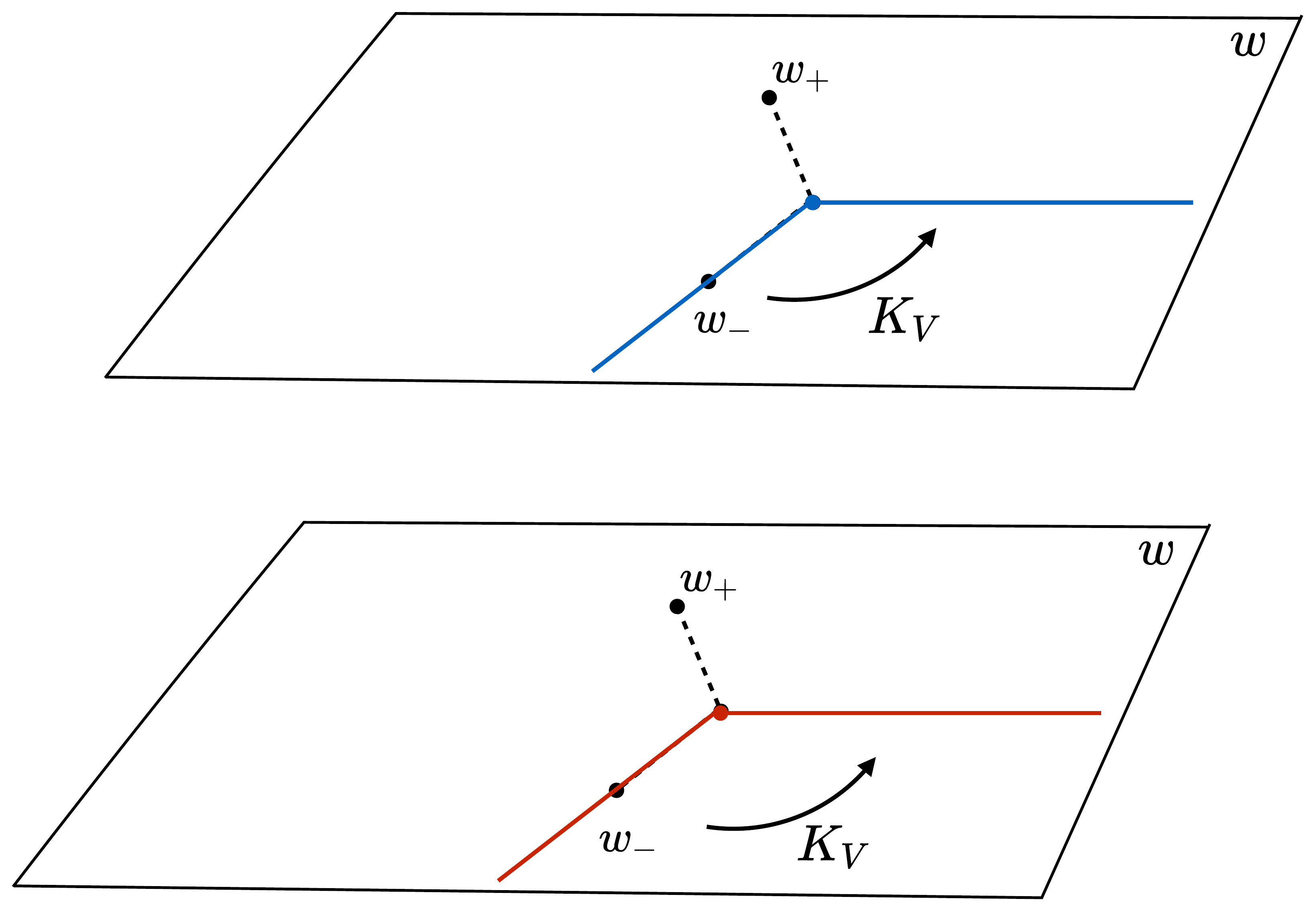}
\caption{The description of the two-interval reduced density matrix in equation \ref{rhoV} is manifest in the $w$ coordinates, where the cutting and re-gluing of the region $V$ under the modular flow is implemented by the branch points of the two-sheeted surface $\Sigma$.  The figures are snap shots of the modular flow and moves in order from left, right to bottom.  For clarity, we have deformed the cut connecting the two branch points relative to the previous figure.}
 \label{figure:C&G}
\end{figure}

We begin with the key observation that the mapping \eqref{w} is single valued provided we treat $w$  like a local coordinate on a Riemann surface $ \Sigma_{2}$, which is a two fold branched cover of the $w$-plane  (right of figure \ref{figure:surface}).  The two branch points in this covering corresponding to $z_{\pm}$ are located at 
\bea
w_{\pm}   = -(\frac{\sqrt{a}\mp i \sqrt{b}}{\sqrt{a} \pm i \sqrt{b}})^2 .
\eea
In this coordinate system, $V$ consists of two copies of the half infinite line $\mathbb{R}^{+}$ living on the two sheets, and the two-interval modular flow is mapped to the simultaneous evolution of these two half lines that sweeps out the surface $\Sigma_{2}$.  This evolution starts off as a simultaneous rigid rotation, but is then interrupted by the presence of the branch point joining the two sheets.   The branch point cuts and re-glues the two half lines as shown in the figure \ref{figure:C&G}.  This is the same cutting and re-gluing process described earlier on the $z$ plane, except that the process is automatically implemented by the topology of $\Sigma_{2}$.

The advantage of working in the $w$ coordinate system is that the effects of the branch points can be represented by the type of twisted boundary conditions that appears in orbifold CFT's.  To do this, we orbifold the surface $\Sigma_{2}$ by identifying points on the two sheets with the same $w$ coordinate. This leads to a theory of two ``replica" fields $ \Psi_{1} (w), \Psi_{2} (w) $ on a  single-sheeted surface with nontrivial monodromies around $w_{\pm} $ : 
\begin{align} \label{mon}
\Psi_{1} (e^{ 2\pi i} w_{+} ) &= \Psi_{2} (w_{+})\nn 
\Psi_{2} ( e^{ 2\pi i} w_{+} ) &= -\Psi_{1} (w_{+})\\
\Psi_{1} (e^{ 2\pi i} w_{-} ) &= -\Psi_{2} (w_{-})\nn
\Psi_{2} ( e^{ 2\pi i} w_{-} ) &= +\Psi_{1} (w_{-})
\end{align} 
Note the minus sign in the second and third equation due to the rotation of the fermion around the branch points.  
Assuming identical boundary conditions for $\Psi_{1}$ and $\Psi_{2}$ at the regulator surfaces near $w=0$ and $w=\infty$, the monodromies can be diagonalized by defining
\bea
\Psi_{\pm} = \frac{1}{\sqrt{2}}  (\Psi_{1} \pm i  \Psi_{2} ),
\eea which satisfy
\begin{align} \label{mon2}
\Psi_{\pm} ( e^{ 2\pi i} w_{+} ) &= \pm i \Psi_{+} (w_{+})\nn
\Psi_{\pm} ( e^{ 2\pi i} w_{-} ) &= \mp i \Psi_{+} (w_{+})
\end{align} 
The effective partition function  $Z_{V}=\tr_{V}  e^{-H_{V}} $ can be express by the path integral 
\begin{align}  \label{EP}
Z_{V} &=\int [D\Psi_{\pm} D \bar{\Psi}_{\pm}]_{\text{twisted}}   e^{-S[\Psi_{\pm}, \bar{\Psi}_{\pm}]  }\nn
S &= \int  \Psi_{+} \bar{\pd} \Psi_{+}  + \Psi_{-} \bar{\pd} \Psi_{-}  d^{2} w +  \int  \bar{\Psi}_{+} \pd \bar{\Psi}_{+}  + \bar{\Psi}_{-} {\pd} \bar{\Psi} _{-}  d^{2} \bar{w} 
\end{align}  
where the twist boundary conditions \eqref{mon} have been imposed in the measure.  To obtain the modular action, we have to incorporate these boundary conditions into the action.  
For free fermions, this can be done by coupling  $ (\Psi_{+},\Psi_{-})$ to a fictitious background U(1) gauge field with holomorphic and anti holomorphic components :
\bea
\mathcal{A}= \begin{pmatrix} A^{+}  &&0\\0&&A^{-}     \end{pmatrix} \quad \quad  \bar{\mathcal{A}}= \begin{pmatrix} \bar{A}^{+} &&0\\0&& \bar{A}^{-} 
  \end{pmatrix}
\eea
Then the monodromies \eqref{mon} can be interpreted as specifying the non-trivial parallel transport of $\Psi_{\pm}$ around $w_{\pm}$.  
 The required gauge fluxes \footnote{Explicitly, the gauge fluxes reproducing \eqref{mon} are
\begin{align}
\int_{C_{+}} A^{+} &=  \frac{\pi}{2}, \quad\quad \int_{C_{-}} A^{+} = - \frac{\pi}{2} \nn
\int_{C_{+}} A^{-} &=  -\frac{\pi}{2},  \quad \quad  \int_{C_{-}} A^{-}  =  \frac{\pi}{2} 
\end{align} 
where $C_{\pm}$ denote counterclockwise loops encircling $w_{\pm}$. }  are realized by the (singular) gauge fields
\begin{align}
A^{+} &= \frac{1}{4i} ( \frac{1}{w-w_{+}}  -  \frac{1}{w-w_{-}})dw,\quad \quad A^{-} = - A^{+}   \nn
\bar{A}^{+} &= -\frac{1}{4i} ( \frac{1}{\bar{w}-\bar{w}_{+}}  -  \frac{1}{\bar{w}-\bar{w}_{-}})d \bar{w},\quad \quad\bar{A}^{-} = - \bar{A}^{+} 
\end{align} 
Instead of imposing twisted boundary conditions in the measure for $Z_{V}$ as in  \eqref{EP}, we couple unconstrained free fermions $\Psi_{\pm}$ to the background gauge field $\mathcal{A}, \bar{\mathcal{A}}$.  This defines the action
\bea
S_{V} = \int_{\Sigma}  \Psi_{+} \bar{\pd} \Psi_{+}  + \Psi_{-} \bar{\pd} \Psi_{-} +i  \bar{A}^{+} (\Psi_{+}^{\dagger} \Psi_{+} -\Psi_{-}^{\dagger} \Psi_{-} ) +\text{Anti-holomorphic part}
\eea
In terms of $S_{V}$, the effective partition function becomes: 
\bea \label{ZV1}
Z_{V} = \int D\Psi_{\pm} D \bar{\Psi}_{\pm}   e^{-S_{V} [\Psi_{\pm}, \bar{\Psi}_{\pm}, A^+, \bar{A}^{+} ]  }
\eea
The path integral measure is now the usual one on the $w$ plane, with no twisted boundary conditions.   Note that despite the singularities in the gauge fields at $w_{\pm}$, the Lagrangian for $S_{V}$  is smooth everywhere because $A^{+}, \bar{A}^+$ couples to the $U(1)$ currents
\begin{align}\label{current}
\mathcal{J}(w)&=  \Psi^{\dagger}_{+}  \Psi_{+} - \Psi^{\dagger}_{-}  \Psi_{-}  =i(\Psi^{\dagger}_{1}  \Psi_{2} - \Psi^{\dagger}_{2}  \Psi_{1})\nn
\mathcal{\bar{J}}\bar{(}w)&=  \bar{\Psi}^{\dagger}_{+}  \bar{\Psi}_{+} - \bar{\Psi}^{\dagger}_{-}  \bar{\Psi}_{-}  =i( \bar{\Psi}^{\dagger}_{1}  \bar{\Psi}_{2} - \bar{\Psi}^{\dagger}_{2}  \bar{\Psi}_{1} )
\end{align}
which vanish at $w_{\pm}$, where $ \Psi_{1} = \Psi_{2}$.  The original problem of obtaining the two -interval modular Hamiltonian is now mapped to problem of obtaining the half line modular Hamiltonian for $Z_{V}$.   

\begin{figure}
\centering
\includegraphics[scale=.20]{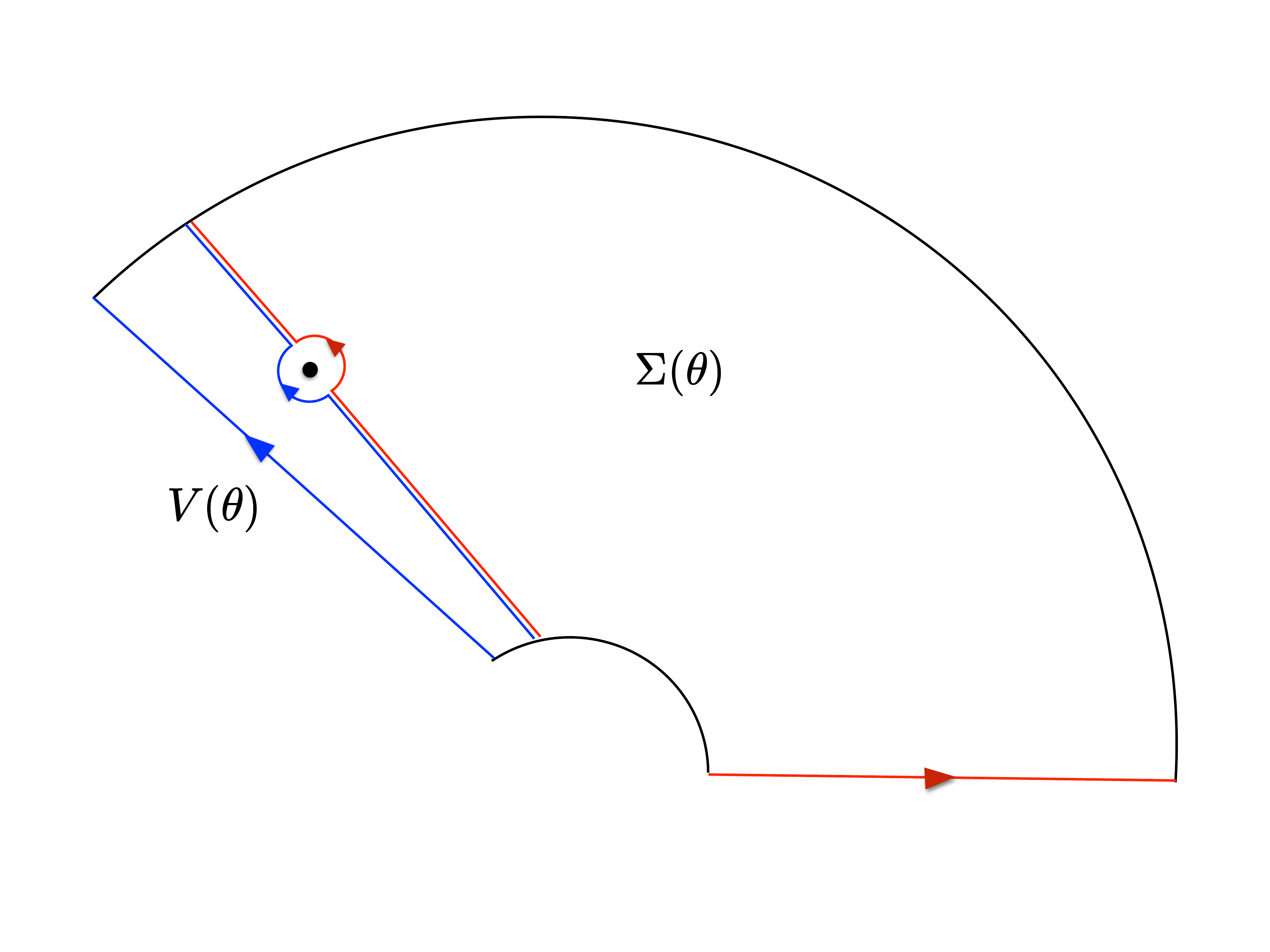}
\caption{The figure shows the modular evolution of a half space by angle $\theta$ on the regulated $w$ plane.  This is generated by the charge $K(\theta)= \int_{V(\theta)} \, w T_{V} (w) d w  +  \int_{V(\theta)} \, \bar{w}  \bar{T}_{V} (\bar{w}) d \bar{w} $, which is integrated along outwardly oriented radial lines $V(\theta)$.  Comparing this operator before (red) and after (blue) the point $w_{\pm}$ shows that it changes by the integral of $ w T_{V} +\bar{w} \bar{T}_{V}  $ along a loop encircling $w_{\pm}$ . }
\label{wedge3}
\end{figure}

Now we come to the crucial point.   The expression \eqref{ZV1} for $Z_{V}$ makes manifest a rotational invariance around the entangling surface $w=0$.  This means that $S_{V}$ defines a finite temperature field theory with the angular direction around $w$ as the temperature circle, and is therefore the modular action we have been searching for.   To show this, we construct the charge that generates rotations around $w=0$ and show that it is explicitly conserved.   This is given by 
\begin{align}\label{K}
K(\theta) = \int_{V(\theta)} \, w T_{V} (w) d w  +  \int_{V(\theta)} \, \bar{w}  \bar{T}_{V} (\bar{w}) d \bar{w} 
\end{align} 
where $ V(\theta)$ is a ray leaving the origin at angle $\theta$, and $T_{V}, \bar{T}_{V}$ are the components of \emph{modular} stress tensor derived from the action $S_{V}$.   In the original gauge,  we would interpret $T_{V},\bar{T}_{V}  $ as the quantum stress tensor containing a contribution path-integral measure.  Note that the expression in \eqref{K}  for the rotational generator assumes a traceless stress tensor, implying that $S_{V}$ also defines a conformal field theory.   We will show this is indeed the case in the next section.    Explicitly, we have
\bea \label{T}
T_{V}&= T  +i A^{+}\mathcal{J}  \nn
\bar{T}_{V} &= \bar{T}  +i \bar{ A}^{+} \bar{\mathcal{J}},
\eea
where $T,\bar{T}$ are the stress tensors for the free fermions $\Psi_{\pm}$ in the absence of the gauge field.  The charge $K$ is conserved provided that $T_{V}$/$\bar{T}_{V}$ are holomorphic/anti-holomorphic ( see figure \ref{wedge3}), which we can check explicitly.
\begin{align}\label{dt}
\bar{\pd} T_{V} (w) &= \frac{1}{4} \mathcal{J}(w)  \bar{\pd} \left( \frac{1}{w-w_{+}}  -   \frac{1}{w-w_{-}} \right)\nn
&= \frac{1}{4} \mathcal{J}(w)  \left(\delta (w-w_{+}) -\delta (w-w_{-} ) \right)=0
\end{align} 
due to the vanishing of the $\mathcal{J}$ at $w_{\pm}$.
Since $K$ is conserved, we can apply \eqref{local} to get the half space modular hamiltonian on the $w$ plane:
\begin{align}\label{Mod}
H_{\mathbb{R}^+} &= \int_{\mathbb{R}^+ } 2 \pi w \, T_{V}(w) \, d w \nn
&=  \int_{\mathbb{R}^+} 2 \pi w  \left(\sum_{a=\pm}  \frac{\Psi^{\dagger}_{a} \pd_{w}\Psi_{a} +  \Psi_{a} \pd_{w}\Psi^{\dagger}_{a}}{2}  + i A^{+} (\Psi^{\dagger}_{+}\Psi_{+} - \Psi^{\dagger}_{-}\Psi_{-} ) \right) \, dw 
 \end{align}
To see the non-locality of this modular Hamiltonian,  we rotate back to the  $\Psi_{1,2}(w)$ basis and observe the coupling: 
\begin{align} \label{bl}
\delta H_{\mathbb{R}^+} &=   
2 \pi i  \int  w  A^{+}\left(  \Psi^{\dagger}_{+}\Psi_{+}  -\Psi^{\dagger}_{-}\Psi_{-}  \right)   \, dw\nn
&=  2 \pi  \int  w \,\Psi^{\dagger}_{i} (w) \mathcal{A}_{ij} \Psi_{j}(w) \, dw 
\end{align}
where 
\bea 
\mathcal{A}_{ij} = \frac{\epsilon_{ij}}{4} \frac{(w_{+} -w_{-} )}{(w-w_{+})(w-w_{-})}
\eea
In the $w$ coordinate system, this looks like a local coupling of two fermion fields.   However, when mapped to the $z$ plane, this will become a \emph{bi-local} coupling of single fermion field evaluated at two different locations.

To make a full comparison to the modular Hamiltonian \eqref{CH}, we now transform back to the $z$ coordinates.
To do so, note that we have applied the mapping \eqref{w} as a \emph{diffeomorphism}, rather than a conformal map.    In particular, the singularities at $w_{\pm}$ are \emph{coordinate} singularities rather than curvature singularities.  This is manifest in the original $z$ plane, since the path integral defining the reduced density matrix $\rho_{V}$ was non-singular at $z_{\pm}$.  Instead, $z_{\pm}$ are singularities of the two-interval foliation, which exactly matches the branch point singularities of the mapping \eqref{w}.    Therefore we do not need to account for the conformal anomaly when we transform back to the $z$ coordinates.

Second, note that the inverse $z=f^{-1} (w)$ is multi-valued, so we must choose a branch. 
In the $z$ coordinates, orbifolding $\Sigma_{2}$ corresponds to identifying the two points \footnote{We can think of $z_{1,2}(w)$ as coordinates on the two different sheets of $\Sigma$} $z_{1}(w)$ and $z_{2}(w)$ that are mapped to the same $w$ under \eqref{w}.  These are two roots of a quadratic polynomial that are related by 
\bea \label{z12}
z_{1} z_{2} = - a b 
\eea  
The orbifold theory thus consists of a two component fermion field $(\Psi(z_{1}), \Psi (z_{2}) )$ living on half of the $z$ plane (see region 1 in figure \ref{regions}); we choose this to be the half containing the left interval $V_{1}= (-b,-a)$ and coordinatized by $z_{1}$.    Given this choice, and treating $z_{2}$ as a function of $z_{1}$ given by \eqref{z12}, we can transform $\delta H_{\mathbb{R}^+} $ according to
\begin{align}
\Psi_{1} (w) &= (\frac{d z_{1}}{d w})^{1/2} \Psi (z_{1} )\nn
\Psi_{2} (w) &= (\frac{d z_{2}}{d w})^{1/2} \Psi (z_{2 } )\nn
d w &=  \frac{d w }{d z_{1}} dz_{1}   
\end{align}
which gives
\begin{align} \label{deltaH}
\delta H_{V} &= -2 \pi    \int_{V_{1} } \left(\frac{ d \log \,  w }{d z_{2} }\right)^{-1}    \frac{  \Psi^{\dagger} (z_{1}) \Psi(z_{2}) -  \Psi^{\dagger} (z_{2}) \Psi(z_{1})}{z_{1}-z_{2}} d z_{1},\nn
&= -2 \pi    \sum_{j\neq i} \int_{V_{i}}  \left(\frac{ d \log \,  w }{d z_{j} }\right)^{-1}   \frac{ \Psi^{\dagger} (z_{i}) \Psi(z_{j})}{z_{i}-z_{j}}  dz_{i}
\end{align} 
Lifting back from the orbifold to the $z$ plane, this term represents a \emph{bi local} coupling of the fields at $z_{1}$ and $z_{2}$.  This is the same bilocal term that appears in equation \eqref{CH} and was first derived in \cite{casini2009reduced}.   Similarly, transforming the kinetic terms $ 2 \pi \int  w\, dw \, \left( \Psi^{\dagger}_{+} \pd \Psi_{+} + \Psi_{-} \pd \Psi^{\dagger}_{-} + \text{h.c.} \right)$ in \eqref{Mod} back to the $z$ plane gives the local term in the modular Hamiltonian \eqref{CH}.
\bea
H_{V}^{\text{local}} =  2 \pi \int_{V}  \left(\frac{ d \log \,  w }{d z}\right)^{-1} \left ( \frac{\Psi^{\dagger} (z) \pd \Psi(z) + \Psi (z) \pd \Psi^{\dagger}(z) }{2}\right) \, dz 
\eea 
Even though we have derived \eqref{deltaH} for $\delta H_{V}$ using a particularly symmetric choice of intervals, the result holds for a general pair of intervals $V= (a_{1},b_{1}) \cup (a_{2},b_{2})$.  Showing this directly is tedious because solving for $z_{2}$ in terms of $z_{1}$ and $a_{i}, b_{i}$ is more complicated in general, since they are the roots of the polynomial 
\bea
p(z)= w \prod^{2}_{i=1}( z-b_{i}) - \prod^{2}_{i=1}( z-a_{i})
\eea  
However one can circumvent this calculation by noting that for generic intervals, pulling back $\delta H_{\mathbb{R}^{+}}$ via the function $z_{1}(w)$  gives
\begin{align}\label{deltaH2}
\delta H_{V} &= \int_{V_{1}}  \left(\frac{d\log w}{d z_{2}}\right)^{-1}  \sqrt{\frac{ d z_{1}}{d z_{2}} }\mathcal{A}_{ij}\Psi^{\dagger}(z_{i}) \Psi(z_{j}) \, d z_{1} \\
\mathcal{A}_{ij} (z_{1}) &= \frac{dw}{ d z_{1}} \mathcal{A}_{ij} (w),\end{align}
where $\mathcal{A}_{ij} (z_{1})$ is the gauge field implementing the monodromies \eqref{mon} in the $z_{1}$ coordinate system.   
For the special case  $V= (-b,-a)\cup(a,b)$, imposing that $\mathcal{A}_{ij} (z_{1})$ has the correct poles and residues at $ z_{\pm}$ gives
\begin{align} \label{A} 
\mathcal{A}_{ij} (z_{1}) &= \frac{\sqrt{ab} }{2} \frac{1}{(z_{1}- i \sqrt{ab})(z_{1}+i \sqrt{ab})} \nn
&= \frac{\epsilon_{ij}}{4} \frac{\sqrt{ \frac{dz_{2}}{dz_{1}} }}{z_{1}-z_{2}},
\end{align} 
where in the second equality we applied  \eqref{z12}.  Moreover, this expression remains valid for arbitrary pairs of intervals, since it makes no reference to the endpoints $(a_{i},b_{i})$.  Plugging this back into \eqref{deltaH2} gives the general result \eqref{deltaH}.

\subsection{Generalization to many intervals and finite size}
The modular Hamiltonian for multiple intervals can be obtained by directly generalizing the orbifold method in section $2.1$.  It will be convenient to consider periodic (Neveu-Schwarz) fermions living on the unit circle in the radially quantized $z$ plane.   The reduced density matrix is given by the path integral on the $z$ plane with cuts along a
set of segments $V= \cup_{i} (a_{i},b_{i})$ of the unit circle.  The change of coordinates
\bea \label{cm}
w=\prod^{n}_{i=1}  \frac{z- a_{i}}{z-b_{i} } 
\eea
takes the $z$ plane to a generically $n$ sheeted surface $\Sigma_{n}$ with local coordinate $w$.  The $n$ sheets are labelled by the roots $z_{i}(w)$ of the polynomial 
\bea
f(w,z)= w \prod^{n}_{i=1} {z-b_{i} } -\prod^{n}_{i=1} z- a_{i}
\eea
and branch points connect multiple sheets where ever  the roots collide. 
To simplify the analysis, we consider the case where intervals are symmetrically placed around the circle, so that for $j\neq 1$
\bea
a_{j}= r^j a_{1}\\
b_{j}= r^j b_{1}\\
r= e^{\frac{2\pi i}{n}}
\eea 
In this case the $n$ roots are related by 
\bea \label{z1}
z_{j}(w) = r^{j} z_{1}(w)
\eea
and there are only two branch points at $z=0,\infty$, each with multiplicity $n$.
On $\Sigma_{n}$, the two branch points are located at $w_{+}=1$ and $w_{-}= e^{-i L}$, where $L$ is the total length of $V$.  As before, we orbifold $\Sigma_{n}$ by identifying $z_{i}(w)$ for $i=1 \cdots n $ to obtain an $n$- component field
\bea
\Psi= (\Psi_{1}(w),\cdots, \Psi_{n}(w))
\eea
with non-trivial holonnomy around $w_{-}$:
\begin{align} \label{T}
\Psi &\rightarrow T \Psi  \nn
T= &\begin{pmatrix}
 0&&1&&0&&\cdots&& 0\\
 \vdots&&0&&1&&0 && \vdots\\ 
 \vdots&&\vdots&&0&&\ddots &&0\\
 \vdots&&\vdots&&\vdots&&\ddots&&1 \\
 (-1)^{n+1}&&0&&\cdots&&0&&0 \\
 \end{pmatrix}
\end{align}
and similarly around $w_{+}$ with holonomy $T^{-1}$.  As noted before, the minus sign in the lower left corner is due to the non-trivial spin of the fermion \cite{casini2005entanglement}.
The holonomy \eqref{T} can be described as a exponential of a matrix gauge field $\mathcal{A}_{ij}$  
\bea
T= \exp i \int_{C_{\pm}}   \mathcal{A} 
\eea
where the $C_{\pm}$ encircle $w_{\pm}$.
For even $n$, $T$ is diagonal in the basis 
\footnote{For the purpose of diagonalizing the twist, we could also have written 
\bea
\Phi_{k}(w)= \frac{1}{\sqrt{n}} \sum_{j} r^{(\frac{1}{2}-k) j } \Psi_{j}(w)
\eea 
with $k=1,\cdots n $, which just permutes the eigenvectors chosen above.  The corresponding transformation matrix $B_{kj}=  r^{(\frac{1}{2}-k) j} $ appears in  in the diagonalization of the multi-interval modular flow.   The eigenvalues obviously doesn't change.   However, we have made a particular chose of phases in the expressions for $t_{k}$, which does affect the gauge fields.  Even though we could in principle shift the phases by integer multiples of $2\pi$ with out change $t_{k}$ , these shifts would change   Modular Hamiltonian and effectively select out a state different than the ground state.  } 

\begin{align} \label{DT}
\Phi_{k}(w)&= \frac{1}{\sqrt{n}} \sum_{j} r^{(k-\frac{n+1}{2}) j } \Psi_{j}(w),\nn
 k&=1,2,\cdots, n
\end{align}
 consisting of eigenvectors with eigenvalue $ t_{k} = e^{\frac{2\pi i }{n} (\frac{n+1}{2}-k)}$.  In this basis, $\mathcal{A}$ has diagonal entries
\bea
A_{k}= \frac{1}{n}(\frac{n+1}{2}-k)\frac{1}{i}\left( \frac{1}{w-w_{+}} - \frac{1}{w-w_{-}}\right)\nn
 k=1,2,\cdots, n
\eea
so that coupling term in the modular Hamiltonian analogous to \eqref{bl} is
\begin{align}
\delta H_{R^+} &=  2 \pi i\int_{R^{+}} w A_{k} \Phi^{\dagger}_{k}(w)\Phi_{k}(w) \, dw 
\end{align}
Undoing the diagonalization then expresses this coupling in terms of an antisymmetric matrix gauge field \cite{rehren2013multilocal}:
\begin{align} 
\mathcal{A}_{ij}&= \frac{1}{i}\left( \frac{1}{w-w_{+}} - \frac{1}{w-w_{-}}\right) \frac{ -r^{\frac{i+j}{2}}}{r^{i} -r^{j}}  \nn
i & \neq j \nn
\delta H_{R^+} &= 2\pi i \int_{R^{+}} w \mathcal{A}_{ij}\Psi^{\dagger}_{i}(w)\Psi_{j}(w) \, dw \nn
\end{align}
Next we pull back to the orbifolded $z$ plane using \eqref{cm}.  
As before it will be illuminating to write down expressions in terms   of the root functions $z_{i}(w)$.  To do so, we first apply conformal map $u= \log w$, which maps the $w$ plane to a cylinder and sends the half line $\mathbb{R}^{+}$ to the entire real line $\mathbb{R}$. Under this mapping 
\begin{align} \label{HR} 
\delta H_{\mathbb{R}^{+}} &\rightarrow  \delta H_{\mathbb{R}}= 2\pi i \int_{R} K_{ij} \Psi^{\dagger}_{i}(u)\Psi_{j}(u) du \\
K_{ij} &= w \mathcal{A}_{ij}(w)  = \frac{dw}{du} \mathcal{A}_{ij}(w) \nonumber
\end{align} 
where $K_{ij}$ is the gauge field transformed into the $u$ coordinates.  For symmetric intervals, it can be expressed in terms of the root functions by using \eqref{z1}:\begin{align}
K_{ij} &= - \frac{\sqrt{z'_{i}(u)} \sqrt{z'_{j}(u)}}{z_{i}(u) -z_{j}(u)}\nn
&=\frac{ -r^{\frac{i+j}{2}}}{r^{i} -r^{j}} \frac{\frac{dz_{i}}{du}}{z_{i}}
\end{align} 
We can verify that this is the correct expression by pulling back to the $z$ plane
\bea
 K_{ij} du = \frac{ -r^{\frac{i+j}{2}}}{r^{i} -r^{j}}  \frac{dz_{1}}{z_{1}}
\eea
and noting that the right hand side has correct poles and residues  at $z_{1}=0,\infty$.  Transforming each term in \eqref{HR} back to the $z_{i}$ coordinates give: 
\bea
\delta H_{V} = -2 \pi   \sum_{j>i} \int_{V_{i}}  \left(\frac{ d u }{d z_{j} }\right)^{-1}   \frac{ \Psi^{\dagger} (z_{i}) \Psi(z_{j})}{z_{i}-z_{j}}  dz_{i}
\eea
This is consistent with the finite size result of \cite{Klich:2015ina} when we make a further conformal mapping $ \zeta= - i \log z_{1}$. 

\section{ Non-local symmetries of $\rho_{V}$}
In this section, we discuss the non-local symmetries of the reduced density matrix, which were first observed in \cite{rehren2013multilocal}.  In particular we will elaborate on the Virasoro symmetry generated by the modular stress tensor $T_{V}$, which played a central role in the derivation of the mult-interval modular Hamiltonian.  Moreover, we will re-interpret our orbifold construction in the original $z$ plane to better understand the boundary conditions which coupled the single-interval modular flows to produce the multi-interval flow.  Finally, we re-derive an isomorphism between the single interval operator algebra of $n$ fermions and the $n$ interval algebra of a single fermion, making a connection to the original observation in  \cite{rehren2013multilocal}.

\subsection{Symmetries of the modular action }
In the previous section we expressed the mult-interval modular hamiltonian in terms of a stress tensor $T_{V}(w)$ associated with the modular action:
\bea
S_{V} = \int \sum_{k=1}^{n} \left( \Phi^{\dagger}_{k} \bar{\pd}\Phi_{k} + i \bar{A}_{k} \Phi^{\dagger}_{k}\Phi_{k} \right) \, d^{2}w  + \text{anti-holomorphic} 
\eea
This action determines a finite temperature path integral $Z_{V}$ describing the reduced density matrix $\rho_{V}$, with the angular coordinate around $\pd V$ taking the role of Euclidean time.  $Z_{V}$  possesses a manifest $U(1)$ symmetry as well as a conformal symmetry that maps to non-local symmetries in the $z$ coordinates due to the non-local nature of the field rotation \eqref{cm}.   

To make this explcit, let's return to the two interval case where \bea \label{SR}
S_{V} = \int  \Psi_{+} \bar{\pd} \Psi_{+}  + \Psi_{-} \bar{\pd} \Psi_{-} +i  \bar{A}^{+} (\Psi_{+}^{\dagger} \Psi_{+} -\Psi_{-}^{\dagger} \Psi_{-} ) d^{2}w+ \text{anti-holomorphic} 
\eea
This possesses the usual $U(1)\times U(1) $ Kacs-Moody symmetry due to the independent, phase rotations of $\Psi_{\pm}(w)$: 
\bea \label{U1}
\Psi_{\pm} (w)  \rightarrow e^{\pm i \theta (w,\bar{w} ) }\Psi_{\pm} (w)\quad \quad A \rightarrow A + \pd \theta \quad\quad \bar{A} \rightarrow  \bar{A} + \bar{\pd} \theta 
\eea
The corresponding conserved currents are  $J_{\pm}(w) =\Psi^{\dagger}_{\pm} \Psi_{\pm}$.   However, it is more illuminating to consider the linear combination
\begin{align}
J(w)&=\frac{J_{+}+J_{-}}{2}  =\Psi^{\dagger}_{1} \Psi_{1} +\Psi^{\dagger}_{2} \Psi_{2} \nn
\mathcal{J}(w)&=\frac{J_{+}-J_{-}}{2} = i ( \Psi^{\dagger}_{1} \Psi_{2} - \Psi^{\dagger}_{2} \Psi_{1}) 
\end{align}
The current $J$ generates the simultaneous phase rotation of $\Psi_{1}$ and $\Psi_{2}$. This is just the $U(1)$ symmetry inherited from the theory on the real line, now restricted to the two intervals. On the other hand $\mathcal{J}$ is a non-local U(1) current that generates a mixing of the fermions in the two intervals.  Moreover,  it is precisely the coupling of this current to it's gauge field $A_{+}(w)$ iwhich connects the two sheets of $\Sigma$, leading to entanglement between the two intervals. 

The action \eqref{SR} is also manifestly invariant under the usual conformal transformation
\begin{align}
w &\rightarrow w + \alpha(w)\nn
\bar{w} &\rightarrow \bar{w} +\bar{\alpha}(\bar{w})\nn
\delta \Psi_{\pm} &=  \alpha \pd \Psi_{\pm} + \frac{1}{2} \pd \alpha \Psi_{\pm} \\
\delta \bar{A}^{+} &= \bar{A}^{+}  \bar{\pd} \bar{\alpha}\label{Abar}  + \bar{\alpha} \bar{\pd} \bar{A}^{+}
\end{align}
for a holomorphic/anti-holomorphic functions $\alpha (w)$/$\bar{\alpha}(\bar{w})$.  However, we must remember that the gauge field $\bar{A}^+$ is not an independent dynamical variable so the one-form transformation \eqref{Abar} must be implemented by applying a gauge transformation to $\Psi_{\pm}$ instead. 

To understand this coupling of gauge and conformal transformation, it's useful to first gauge away the gauge fields from the modular action, so that \eqref{SR} is modified to
\begin{align} 
S&= \int  \Psi'_{+} \bar{\pd} \Psi'_{+}  + \Psi'_{-} \bar{\pd} \Psi'_{-} \,\, d^{2} w + \text{anti-holomorphic}  \nn
\Psi'_{\pm} &= \exp \left( i \int^{w} A^{\pm} +i \int^{\bar{w}} \bar{A}^{\pm}\right)  \Psi_{\pm}
\end{align}
where $\Psi_{\pm}$ have trivial monodromy.  
Now when we apply a conformal transformation to $\Psi'_{\pm}$ we must explicitly transform the phase factor: 

\begin{align} 
\Psi'_{\pm}(w+ \alpha, \bar{w}+\bar{\alpha}) &= \exp ( i \int^{w+\alpha} A^{\pm} +i \int^{\bar{w}+\bar{\alpha} } \bar{A}^{\pm}) \Psi_{\pm}(w+\alpha,\bar{w} + \bar{\alpha} )\nn
&\sim\exp ( i \alpha A^{\pm} +i \bar{\alpha}  \bar{A}^{\pm}) \Psi'_{\pm}(w+\alpha,\bar{w} + \bar{\alpha} )
\end{align}
This is a field dependent gauge transformation that shifts the gauge field by a variation appropriate to a conformal transformation: 
\bea
A\rightarrow A +  \pd \alpha A + \alpha \pd A \nn
\bar{A}\rightarrow  \bar{A} +  \bar{\pd} \bar{\alpha} \bar{A} +\bar{\alpha} \bar{\pd} \bar{ A}
\eea
This leads an additional variation of the field under conformal maps:
\bea \label{nlr}
\delta_{\text{nl}} \Psi'_{\pm}= i  \left(\alpha A^{\pm} +   \bar{\alpha} \bar{A}^{\pm} \right)  \Psi'_{\pm}
\eea
So the total conformal transformation is
\begin{align} \label{nlc}
w &\rightarrow w + \alpha(w)\nn
\bar{w} &\rightarrow \bar{w} +\bar{\alpha}(\bar{w})\nn
\delta \Psi'_{\pm} &=  \alpha \pd \Psi'_{\pm} +  \frac{1}{2}\pd \alpha \Psi'_{\pm}  +i  \left(\alpha A^{\pm} +  \bar{\alpha} \bar{A}^{\pm} \right)  \Psi'_{\pm}
\end{align}
This conformal transformation is generated by the modular stress tensor $T_{V}$ , as one can check via the Ward identities.  The gauge-field dependent part of this transformation is non-local because it applies an infinitesmal phase rotation to $\Psi_{\pm}$ which mixes $\Psi_{1}$ and $\Psi_{2}$.   It is generated by the part of $ T_{V}(w)$ that is proportional to the U(1) current $\mathcal{J}$.  Finally, note that the central charge of this non-local CFT is twice that of the original free fermion theory, since there are effectively two species of fermions corresponding to the two regions.  

\subsection{Gluing modular flows in the $z$ plane}
\begin{figure}
\centering
\includegraphics[scale=.35]{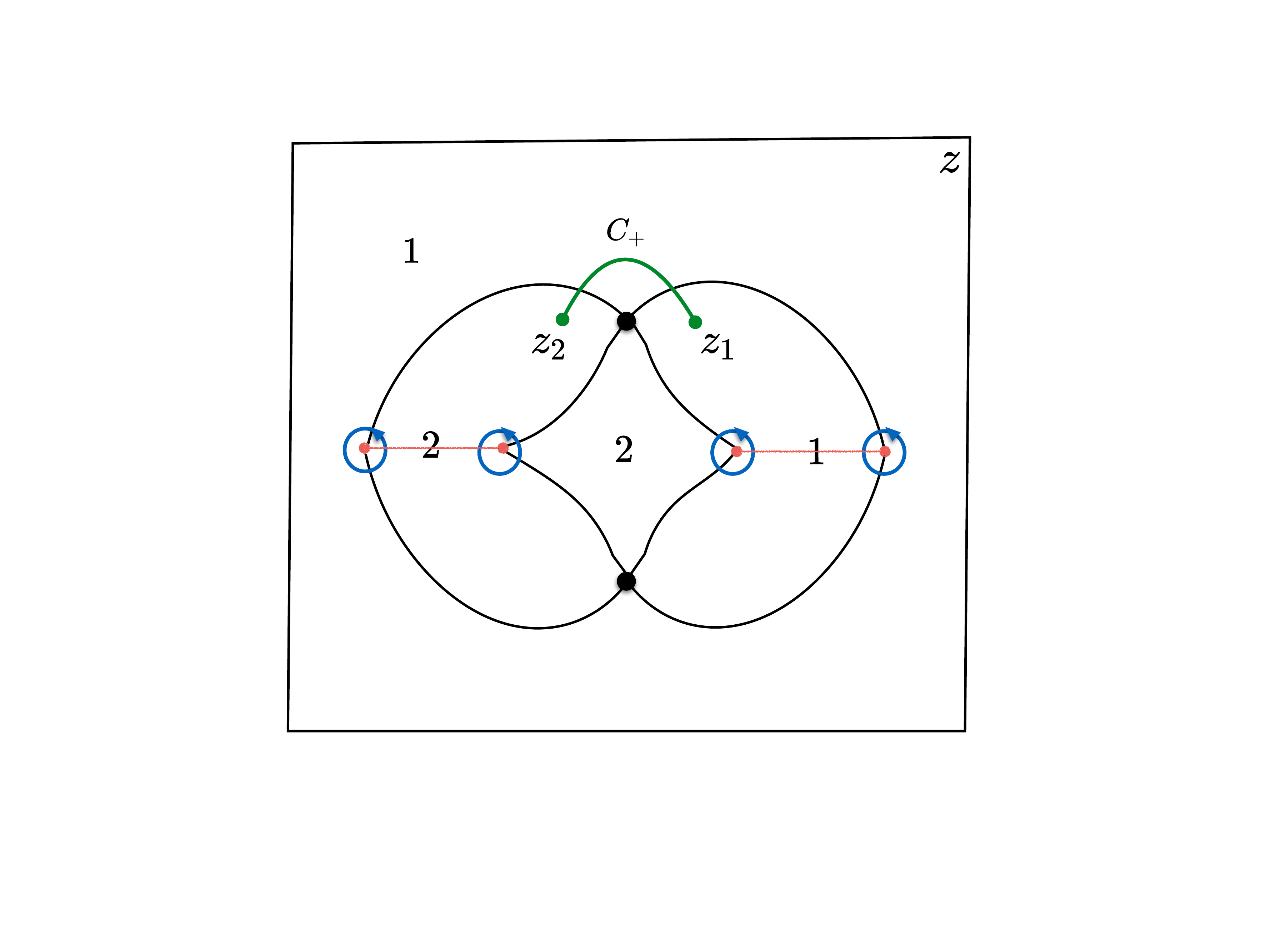}
\caption{The root functions. $z_{1,2}(w)$ define two regions of the complex $z$ plane minus the points $z_{\pm}$.  A closed curve $C_{+}$ around $w_{+}$  on the $w$ plane corresponds to an open curve on the $z$ plane connecting $z_{1}$ to $z_{2}$. }
\label{regions}
\end{figure}

Our derivation of the non-local symmetries of $\rho_{V} $ and  modular Hamiltonian was done entirely in the $w$ coordinate system.  However, it will be insightful to interpret our construction in the $z$ coordinate system via \eqref{w}, where we can see how the single interval modular flows are glued together to form the multi-interval flow.   To begin, recall that the  mapping \eqref{w} for the two-interval case effectively partitions the $z$ plane into two regions defined by the root functions $z_{1,2}(w)$ \footnote{This is really a partition of the twice punctured $z$ plane with $z_{\pm}$  removed.}.  These regions are shown in figure \ref{regions} , and corresponds to the two sheets of $\Sigma_{2}$ in the $w$ coordinate system.   Moreover, there is a $Z_{2} $ symmetry that exchanges  region 1 and 2, corresponding to exchanging the two sheets of $\Sigma_{2}$.

Now consider a free fermion theory describing the two decoupled regions of the $z$ plane.  This is defined by the action:
\begin{align}
S= \int_{\color{red} 1}  \Psi^{\dagger}(z_{1} )\bar{ \partial}_{\color{red} 1} \Psi(z_{1}) d^{2}z_{1}  + \int_{\color{red}2}  \Psi^{\dagger}(z_{2} )\bar{ \partial}_{\color{red} 2}\Psi(z_{2})  d^{2}z_{2}
\end{align} 
Now we orbifold the $z$ plane by pulling back the fields on region 2 to region 1, using the relation $z_{2}= -\frac{ab}{z_{1}} $ between the two roots.   This is a holomorphic transformation from $z_{2}$ to $z_{1}$, under which:
\begin{align}
S= \int_{\color{red} 1}  \left( \Psi^{\dagger}(z_{1} )\bar{ \partial}_{\color{red} 1} \Psi(z_{1})  + \sqrt{\frac{dz_{2}}{dz_{1}}}\Psi^{\dagger}(z_{2} )\bar{ \partial}_{\color{red} 1} \sqrt{\frac{dz_{2}}{dz_{1}}}  \Psi(z_{2}) \right) d^{2}z_{1}
\end{align} 
In the $z_{1}$ coordinates, we can see a manifest non-local $Z_{2}$ symmetry: 
\begin{align}
\Psi (z_{1}) \rightarrow. \sqrt{\frac{dz_{2}}{dz_{1}} } \Psi(z_{2})  \quad\quad \sqrt{\frac{dz_{2}}{dz_{1}}} \Psi(z_{2})  \rightarrow -\Psi(z_{1}) 
\end{align} 
This symmetry is always present for a conformal field theory, because the transformation $z_{2}(z_{1})$ is conformal.   For $n$ intervals, we would have a $Z_{n}$ symmetry that permutes the $n$ roots.  
In the $w$ coordinate system, we glued together the different sheets of $\Sigma_{2}$ by the monodromies \eqref{mon} around  $w_{\pm}$, which are defined by the non-local $Z_{2}$ symmetry.   On the $z$ plane, a closed loop around $w_{\pm}$ is actually an open curve that connects a point $z_{1}$ in region 1 with it's image $z_{2}(z_{1})$ in region 2.   The mondromies \eqref{mon}  are therefore gluing conditions telling us how to match the field in region 1 and region 2 when we orbifold the $z$ plane.   This leads to a gluing of the single interval modular flows describing each individual region into the multi-interval modular flow for the entire Euclidean spacetime. 

For free fermions, this discrete, non-local $Z_{n}$ symmetry is embedded in a non-local $U(1)$ symmetry, which is also a crucial part of the non-local conformal symmetry.   This is generated by a non-local current on the $z$ plane, which we can derive by pulling back the current \eqref{current} from the $w$ plane.  Usually, the transformation of a chiral current is
\bea
J(w)= \frac{dz}{dw} J(z)
\eea
consistent with the invariant one form $J=J(z) dz$. However, in this case we have the transformation law 
\bea
\mathcal{J}(w)= \sqrt{\frac{dz_{1}}{dw} }\sqrt{ \frac{dz_{2}}{dw} } \mathcal{J}(z)
\eea
which is appropriate for the invariant form  $\mathcal{J}= \mathcal{J}(z_{1},z_{2}) \sqrt{dz_{1}}\sqrt{dz_{2}} $, where
\bea
\mathcal{J}(z_{1},z_{2}) = i ( \Psi^{\dagger}(z_{1}) \Psi(z_{2}) - \Psi^{\dagger}(z_{2}) \Psi(z{_1}) ) 
\eea

Similarly, the gauge field $\bar{\mathcal{A}}$ coupling to $\mathcal{J}$ is most elegantly expressed in terms of a bi-local one form in the $z$ coordinates.   From equation \eqref{A} we have:
\begin{align}
\mathcal{\bar{A}} = \mathcal{\bar{A}} (\bar{z}_{1}) d\bar{z}_{1}  &= \frac{1}{4} \frac{\sqrt{ \frac{d\bar{z}_{2}}{d\bar{z}_{1}} }}{\bar{z}_{1}-\bar{z}_{2}} dz_{1}\nn
&\sim \frac{1}{4} \frac{\sqrt{ d\bar{z}_{2} } \sqrt{d\bar{z}_{1}} }{\bar{z}_{1}-\bar{z}_{2}}
\end{align}
Here we have used the $\sim$ symbol to denote equality when pulled back along the equation $\bar{z}_{1}\bar{z}_{2}= -ab$.  This suggest we should define the bi-local gauge field to have anti-holomorphic component 
\bea
\bar{\mathcal{A}}(\bar{z}_{1},\bar{z}_{2}) = \frac{1}{4}\frac{1}{\bar{z}_{1} - \bar{z}_{2}} 
\eea

In terms of these objects the action $S_{V}$ in the $z$ coordinates can be expressed in a way that is symmetric in $z_{1}$ and $z_{2}$. 
\bea \label{NLA}
S_{V} = \int 
\sum_{j=1}^{2} \Psi^{\dagger}(z_{j}) \bar{\pd}  \Psi_{j}(z_{j})   d^{2}z_{j} +
\int   \bar{A}(\bar{z}_{1},\bar{z}_{2}) \mathcal{J}(z_{1},z_{2})  \,\,\sqrt{d z_{1}} \sqrt{d z_{2}}   \wedge \sqrt{d \bar{z}_{1}} \sqrt{d \bar{z}_{2}}
\eea
The $U(1)$ symmetry of this action is generated by the Ward identities 
\begin{align}  \label{Ward}
 \delta \Psi (z_{1}) &=  \int_{C'_{1}} \theta (w) \mathcal{J}'(z'_{1},z'_{2}) \sqrt{\frac{ dz_{2}'}{dz'_{1}}} \Psi(z_{1})\,   \frac{dz_{1}'}{2 \pi i} \nn
 \delta \Psi (z_{2}) &=  \int_{C'_{2}}   \theta (w) \mathcal{J}'(z'_{1},z'_{2}) \sqrt{\frac{ dz_{1}'}{dz'_{2}}} \Psi(z_{1}) \frac{dz_{2}'}{2 \pi i}  \nn
 z_{1}z_{2} &= - ab
\end{align} 
where $\mathcal{J}$ should be pulled back along $ z_{2}(z_{1})$ or $z_{1}(z_{2})$ depending on the location where we vary the field.  The finite transformation is 
\bea 
\begin{pmatrix} 
\Psi(z_{1}) \\ \sqrt{\frac{d z_{2}}{d z_{1}}} \Psi(z_{2})
\end{pmatrix}   \rightarrow \begin{pmatrix} 
\cos (\theta (w)) & \sin (\theta (w)) \\
-\sin (\theta (w)) & \cos (\theta (w)) 
\end{pmatrix} 
\begin{pmatrix} 
\Psi(z_{1}) \\  \sqrt{\frac{dz_{2}}{d z_{1}}} \Psi(z_{2})
\end{pmatrix} 
\eea
As with the $Z_{n}$ symmetry, this symmetry becomes manifest in the action when we transform the measure $d^{2}z_{2}$ via the relation $z_{2} = \frac{-ab}{z_1}$.
Similarly, the non-local conformal symmetry of \eqref{NLA} is generated by the bi-local stress tensor: 
\begin{align}
T_{V}&= T_{V} (z_{1},z_{2}) dz_{1} dz_{2}\nn
T_{V}(z_{1},z_{2})&= \frac{\Psi^{\dagger} (z_{1}) \pd \Psi(z_{1}) +  \Psi^{\dagger} (z_{2}) \pd \Psi(z_{2})}{2} + \text{h.c.}  + i  \frac{  \Psi^{\dagger} (z_{1}) \Psi(z_{2}) -  \Psi^{\dagger} (z_{2}) \Psi(z_{1})}{4(z_{1}-z_{2})} \nn
\end{align}

As noted in  \cite{rehren2013multilocal} this stress tensor gives a non-local  representation of the $c=2$ Virasoro algebra on the two-interval algebra of a single fermion. 

\subsection{Non-local isomorphism of the multi interval fermion algebra}

Our orbifold procedure consists of  applying the singular diffeomorphism  \eqref{cm}
\bea 
w=\prod^{n}_{i=1}  \frac{z- a_{i}}{z-b_{i} } ,
\eea
identifying $z_{i}(w)$ for $i=1,\cdots, n$, and finally performing an internal rotation of the replica fields.  The combined, non-local transformation is:
\begin{align} \label{nl}
\Phi_{k}(w)&= \frac{1}{\sqrt{n}} \sum_{j} r^{(k-\frac{n+1}{2}) j }  \sqrt{\frac{dz_{j}}{dw}} \Psi(z_{j}),\nn
 k&=1,2,\cdots, n
\end{align} 
The fields $\Phi_{k}(w)$ describe $n$ fermions  coupled to a  background gauge field $A_{k}$.  In the $z$ coordinate system, the effect of the gauge fields is to insert a flux at the origin, which modifies the two point function of $\Phi_{k}(w)$ from the free value by the position dependent phase $e^{i \int A_{k}}$.   
Multiplying by the inverse phase gives $n$ \emph{free} fermions
$\tilde{\Phi}_{k}$:
\begin{align}\label{nl}
\tilde{\Phi}_{k}(w)&=e^{-i\int^{w} A_{k}}\Phi_{k}(w)
\nn &= e^{-i\int^{w(z)} A_{k}} \left(\sum_{j} \frac{r^{(k-\frac{n+1}{2}) j }}{\sqrt{n}}  \sqrt{\frac{dz_{j}}{dw}} \Psi(z_{j})\right)\\
 k &=1,2,\cdots, n \nonumber
\end{align}
These fermions have the expectation value:
\bea\label{tp}
\braket {\tilde{\Phi}^{\dagger}_{k}(w)\tilde{\Phi}_{k}(v)} = \frac{1}{w-v}
\eea
Note that this expectation value is a path integral average over $\Phi_{k}$. 
Restricted to the subregion  $V$, equation \eqref{nl} provides a mapping between the single fermion algebra of $\Psi(z)$ on $n$ correlated intervals $V$ and the algebra of $n$ decoupled fermions $\tilde{\Phi}_{k}(w)$ living on a half line $\mathbb{R}^+$.   Due to Wick's theorem, equation \eqref{tp} implies the mapping preserves all expectation values in $V$, while the decoupling between different $\Phi_{k}$'s imply that the mapping preserves all anti-commutators.  Therefore the transformation \eqref{nl} represents a non local isomorphism between the two algebras, as first noted in \cite{rehren2013multilocal}. In this work we have uncovered a \emph{geometric} description of this isomorphism in terms of an orbifolding procedure that glues together the half line modular flows. In the operator language, this gluing corresponds to the inverse of equation \eqref{nl}, in which a U(1) monodromy is imposed on $n$ decoupled fermions with a singular gauge transformation, followed by field rotation that un-diagonalize the monodromy \eqref{mon} and then lifting to a covering space.  This process leads to non trivial correlations and entanglement between the $n$ intervals supporting that fermions $\Psi_{i},\, i=1\dots ,n$.


\subsection{ Explicit comparison with the isomorphism in \cite{rehren2013multilocal}}
In this section we make an explicit comparison between \eqref{nl} and the non- local isomorphism presented in \cite{rehren2013multilocal}.   The authors of \cite{rehren2013multilocal} considered a real chiral fermion $\psi$ in 1+1 D in light cone coordinate $z = x-t$ , which we can identify with our Euclidean coordinate $z=x+i y$ after a wick rotation $ t= -iy $.  Furthermore they transformed to the Cayley coordinate $Z \in S^{1}$, obtained by a stereographical projection of the light cone coordinate onto a unit circle.  Explicitly 
\bea
Z= \frac{1+ i z}{1-i z}
\eea

The stereographic projection is a conformal map, and the real fermion transforms as 
\begin{align} 
\psi(Z)= \sqrt{\frac{dz}{dZ}} \psi(z)=  \frac{1-i z}{\sqrt{2}} \psi(z)
\end{align}
Due to the particular form of the conformal factor,  this field satisfies
\bea \label{*}
\psi(Z)^*= Z\psi(Z)
\eea

In terms of the Cayley variables, \cite{rehren2013multilocal} consider a 2 to 1 conformal mapping $ W=Z^2$, and showed that this induces the isomorphism 
\begin{align} \label{iso}
\phi(W) =\frac{1}{2} ( \psi(Z)  + \psi(-Z) )\nn
\phi^{*}(W) =\frac{1}{2 Z} ( \psi(Z)  + \psi(-Z) )
\end{align}
between a real and complex fermion $\phi$.
We can arrive at the same isomorphism by applying \eqref{nl} for two intervals.   In the Cayley coordinates the branch point is at $Z=0$, so the gauge field with the appropriate holonomy is  :
\begin{align}
A&= \frac{1}{4 i Z}\nn
\bar{A}&=\frac{1}{4 i \bar{Z}}  
\end{align}
Then applying \eqref{nl} gives: 
\begin{align} 
\tilde{\Phi}(W) &= \frac{\exp{ -\frac{1}{4} ( \log Z + \log \bar{Z} )}}{\sqrt{2}} \left( (\frac{d Z_{1}}{dw})^{1/2} \psi(Z_{1}) + i  (\frac{d Z_{2}}{dW})^{1/2} \psi(Z_{2}) \right)\nn
&=\frac{1}{2 \sqrt{Z}}\left( \frac{1}{\sqrt{Z}} \psi(Z) + i  (\frac{1}{\sqrt{-Z}})\psi(-Z)\right)\nn
&=\frac{1}{2Z}\left(\psi(Z) -  \psi(-Z)\right)\nn
\tilde{\Phi}^{\dagger}(W) &= \frac{1}{2}\left(\psi(Z) +  \psi(-Z)\right)
\end{align}
where in the second line I have evaluated on a $t=0$ slice so $Z= \bar{Z}$, and in the last line I used \eqref{*}. This  agrees with \eqref{iso} if we identify
\bea
\tilde{\Phi} \rightarrow  \phi^{*} \nn
\tilde{\Phi}^{\dagger} \rightarrow \phi
\eea
\section{Entanglement entropy from a non-local conformal map} 
We now apply the non-local conformal symmetry of $Z_{V}$ to compute the entanglement entropy of multiple intervals.  As in the single interval case, the strategy is to identify the entanglement entropy with the thermal entropy of the reduced theory represented by $Z_{V}$.  In the case of a single interval, the reduced theory is a thermal BCFT whose entropy is easily computed by via a conformal map to a infinitely long cylinder.   Upon applying a modular transformation, the corresponding partition function and it's entropy can be evaluated as an amplitude between boundary states \cite{Cardy:2016yq}.  For multiple intervals, applying our orbifolding procedure reduces the multiple interval problem to the single interval one, except that the required conformal map is now non-local.   The invariance of the free fermion modular action under such non-local conformal maps will be the crucial ingredient allowing for the derivation of the entanglement entropy (EE).  We will show that the EE can be obtained by integrating a local thermal entropy density, which defines a local entanglement temperature as in the case of a single interval. 

\subsection{Entanglement entropy for single interval }
To warm up, let's review the computation in the case of a single interval $A=(a,b) $, closely following  \cite{Cardy:2016yq}.   The normalized reduced density matrix for the vacuum state has the form
\bea
\rho_{A} = \frac{e^{-H_{A} }}{Z_{A}}
\eea
where $\braket{\phi_{f}|e^{-H_{A}}|\phi_{i}}$ is an unnormalized path integral on the $z$ plane that is cut open along region $A$ and regulated by removing $\epsilon $ sized disks containing $\pd A$.  The regularization introduces a choice of boundary conditions on two small circles around $\pd A$ and defines the effective partition function
\bea
Z_{A} (\epsilon)= \tr_{A} e^{-H_{A}}
\eea
which normalizes the reduced density matrix.  To preserve conformal symmetry we choose conformally invariant boundary conditions at $\pd A$, and $Z_{A}(\epsilon)$ should be interpreted as the partition function of the corresponding  \emph{Boundary} CFT \cite{Cardy:2016yq}.  Explicitly, $Z_{A}(\epsilon)$ is the path integral on a spacetime with the topology of an annulus.
The entanglement entropy can be evaluated directly from $\rho_{A}$:
\bea \label{SA}
S_{A}=\tr( \rho_{A} \log \rho_{A}) = 2 \pi \tr(\rho_{A} K_{A})  + \log Z_{A}
\eea
where we wrote $H_{A}=2 \pi K_{A}$ for later convenience.   To proceed\footnote{The entropy is invariant under conformal transformations even through both the modular energy and free energy terms transform non-trivially due to the conformal anomaly.} we make a conformal transformation to the cylinder geometry parametrized by $u$:
\begin{align}
u(z) &= \log (- \frac{z-a}{z-b})\nn
u &\sim u +  2 \pi  i \nn
\text{Re} \, u &\in ( -\log\frac{b-a}{\epsilon} , \log\frac{b-a}{\epsilon} )
\end{align} 
\begin{figure}
\centering 
\includegraphics[scale=.35]{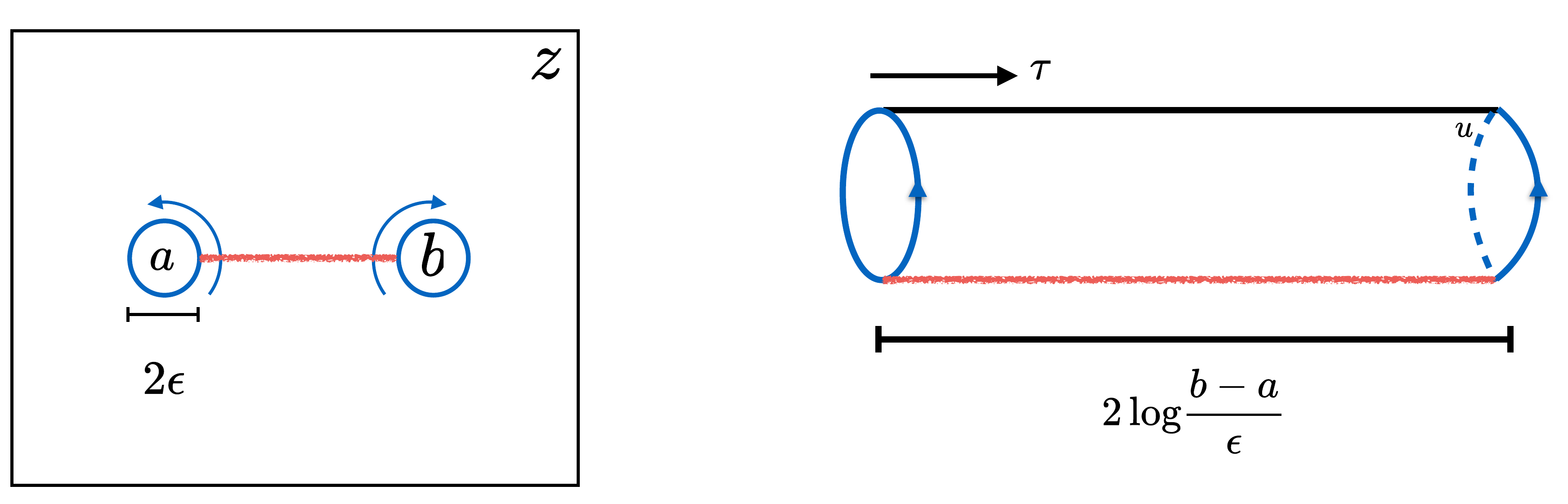}
\caption{The regulated $z$ plane can be conformally mapped to the cylinder geometry on which the modular hamiltonian is equal to the physical hamiltonian of the CFT.  In the ``open string'' channel, this is a thermal state at temperature $2\pi$.   In the ``closed string'' channel, this an amplitude between closed string boundary states living on the Hilbert space of a circle. }
\label{figure:cyl}
\end{figure}
The generator of modular flow $K_{A}$ is mapped to the physical Hamiltonian generating translations around the cylinder , which we denote by 
\bea
U K_{A} U^{-1} = \int_{-\log\frac{b-a}{\epsilon} }^{ \log\frac{b-a}{\epsilon}} T(u)\,  du
\eea
and the conformally transformed density matrix $ \rho= U \rho_{A} U^{-1}$
is thermal with respect to this Hamiltonian at temperature $2\pi$.  Thus,the entanglement entropy is equal to the thermal entropy of this ensemble, and equation \eqref{SA} becomes the standard relation
\bea \label{TEE}
S_{A}= E-F
\eea
between the thermal entropy $S_{A}$, energy $E$, and free energy $F=-\log Z$ of a BCFT at temperature $2\pi$. 
To compute the free energy in the small $\epsilon$ limit, we treat $\Re \, u$ as the Euclidean time variable and interpret $Z$ as an amplitude between boundary states $\ket{a}, \ket{b} $ inserted at the two regulated boundaries of the cylinder.  Letting $ u= \tau +i x $, and $H=L_{0} +\bar{L}_{0} - \frac{c}{12}$ the hamiltonian generating translations in $\tau$, we have 
\begin{align}\label{ZA}
Z &= \braket{a | \exp (-\int_{-\log \frac{b-a}{\epsilon}}^{\log\frac{b-a}{\epsilon}} d \tau    H(\tau) |b}
\nn
&=  \braket{a | \exp ( -2 \log \frac{b-a}{\epsilon} ( L_{0} +\bar{L}_{0} - \frac{c}{12} ) )|b},
\end{align} 
where in the second line we use the fact that system is translation invariant in $\tau$. The boundary states $\ket{a}$ and $ \ket{b}$ can be expanded in terms of Ishibashi states $ \ket{h}\rangle$ which diagonalizes $L_{0}$.   In simple cases, they are in one to one correspondence with primary fields of the bulk CFT and are labelled with corresponding conformal dimensions $h$:
\begin{align}
\ket{h}\rangle&=  \sum_{N} \sum^{d_{h}(N)} _{j=1}  \ket{h; N ; j}_{L} \otimes \ket{\bar{h} ;\bar{N} ; \bar{j}}_{R}\nn
L_{0}\ket{h}\rangle =\bar{L}_{0} \ket{h} \rangle &= (h+N)\ket{h}\rangle
\end{align}
where $L$ and $R$ denote the left and right moving sectors and $d_{h}(N)$ counts the degeneracy at level $N$.   

Expanding \eqref{ZA} in terms of the Ishibashi states, and taking the limit $\epsilon \rightarrow 0$ shows that $Z$ is dominated by the term with $h=N=0$ , which is the vacuum state \emph{on the circular} time slice.  Thus one finds that 
\begin{align}
\lim_{\epsilon \rightarrow 0} \log Z&=  \log  \braket{a|0} \braket{0 | \exp ( -2 \log \frac{b-a}{\epsilon}  (- \frac{c}{12} ) )|0}\braket{0|b}  \nn
&= \frac{1}{6} \log \frac{b-a}{\epsilon}.
\end{align} 
Here we have neglected the subleading contributions due the the boundary entropies $\log  \braket{a|0} $ and $ \log \braket{0|b} $.  This gives the standard CFT result for the free energy on a cylinder, which is just proportional to the Casimir energy.  The corresponding average energy $E$ is
\bea
E =   \frac{1}{6}  \log\frac{b-a}{\epsilon}
\eea 
Adding the two term gives the expected EE
\bea
S_{A}= \frac{1}{3} \log \frac{b-a}{\epsilon}
\eea
As noted in \cite{Wong:2013fk} we could have calculated the EE on the $z$ plane by treating 
$ \beta(z)= \frac{2\pi}{u'(z)}$ as an effective spatially varying temperature.   The EE can then be obtained by integrating the corresponding effective entropy density for a CFT,
\bea \label{int}
S_{A} = \int_{a}^{b}  \frac{c \pi}{3\beta }  dz 
\eea
This integral contains a UV divergence localized to the entangling surface where the entanglement temperature diverges.  In the replica trick, these are locations of curvature singularities due to branch points of the replicated manifold.  On the other hand, the conformal map to the $u$ cylinder maps this UV divergence to an IR divergence: The entanglement entropy is just proportional the casimir energy times \footnote{ In the replica trick, this casimir energy cancels out in the ratio $\frac{\tr_{A}(\rho^{n}_{A})}{\tr_{A}(\rho_{A})^{n}}$, whereas in our thermal calculation it is entirely responsible for the EE! } the spatial volume, which is regulated due to the ``stretching'' of the entangling surface introduced by the regulator $\epsilon$.  We will see below that the multi-interval EE can be calculated in a similar fashion.

\subsection{Entanglement entropy of multiple intervals}
To keep the notation light, let's start with two intervals $V= (a_{1},b_{1})\cup (a_{2},b_{2}) $  on the $z$ plane.  The associated reduced density matrix is defined with conformal boundary conditions at the regulator surfaces near $a_{i}$ and $ b_{i} $.  We have shown that this can be presented as an orbifold theory on the $w$ plane via the diffeomorphism
\bea\label{w'}
w= \frac{(z-a_{1})(z-a_{2})}{(z-b_{1})(z-b_{2})}
\eea 
\begin{figure}
\centering 
\includegraphics[scale=.35]{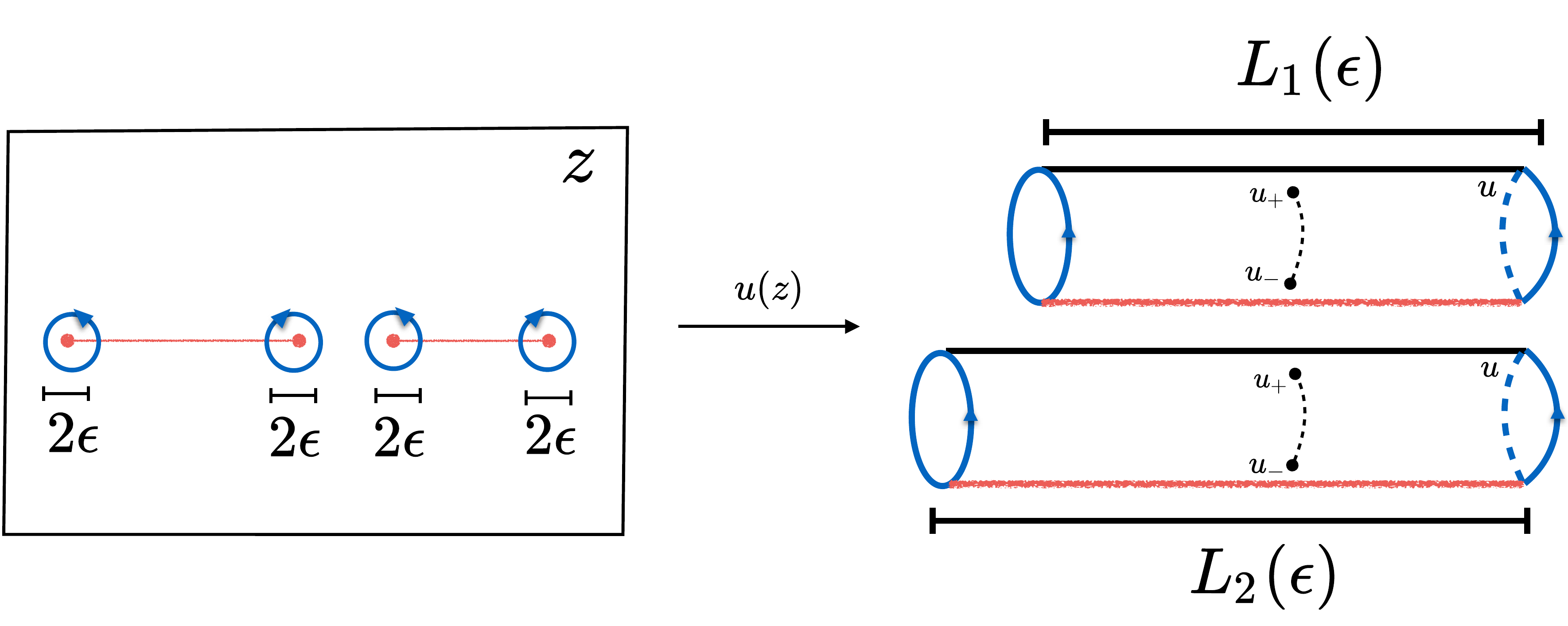}
\caption{The regulated $z$ plane is mapped to a two sheeted cover of the cylinder, with slightly different lengths $L_{1}(\epsilon) $ and  $L_{2}(\epsilon)$.  The total length of the two sheeted cylinder is proportional to the entanglement entropy.   }
\label{cyl2}
\end{figure}
The orbifold theory itself is a conformal field theory of two fields $\Psi_{1,2} (w)$ coupled to a background gauge field $A_{+}$.  We now make a conformal map
\bea \label{u2}
u= \log w
\eea
taking the regulated $w$ plane to the cylinder $\mathcal{C}$.   As noted before, the representation of this conformal map on the original field $\Psi(z)$ is non-local, unlike the single interval case. 
Furthermore, we emphasize that unlike \eqref{w'}, \eqref{u2} is accompanied by a change of the metric by a conformal factor, which implies we should include the Schwarzian when transforming the stress tensor.   
Under the mapping \eqref{u2}, the reduced density matrix $\rho_{V}$ is mapped into a thermal state with respect to the ``physical" Hamiltonian $H$ generating evolution along the thermal circle on the cylinder:
\begin{align}
H&= 2\pi  \int_{-\infty}^{\infty}  T_{V}(u) - \frac{c_{V}}{12}\, du 
\end{align}
In writing the energy density on the cylinder, we have carefully separated the casimir energy density arising from the conformal anomaly.  Note that $c_{V}=2$ is the central charge of the nonlocal CFT, defined by the modular stress tensor $T_{V}$.   The integration here is along real $u$ direction and we have not regulated the infinite volume yet due to a subtlety we will discuss below.   Once we have accounted for the central charge of the non-local CFT, the calculation formally proceeds in the same way as the single interval case.
The entanglement entropy can again be cast as a thermal entropy as in \eqref{TEE}.  After a modular transformation, the free energy can be computed as an amplitude between two boundary states living in the Hilbert space $\mathcal{H}_{S^{1}}$ of a circle at at $\tau = \pm \infty$.  
\begin{align}\label{ZV}
Z_{V} =\braket{a | \exp (-\int d \tau \left( L_{0}+\bar{L}_{0}  - \frac{c_{V}}{12}\right) |b},
\end{align} 
where the Virasoro generators $L_{n}$ defining the ``closed string" Hamiltonian as well as the boundary states $\ket{a}, \ket{b} $ correspond to the non-local conformal symmetry with $c_{V}=2$.   
The long time evolution again projects onto the vacuum state, giving 
\begin{align}
Z_{V} \sim  \exp ( \frac{c_{V}}{12} \int_{-\infty}^{\infty} d \tau)
\end{align} 
Thus we have arrived a leading order free energy that is equal to the casimir energy times the spatial volume of the orbifold CFT with central charge $c_{V}=2$.  This is IR divergent and is regulated by introducing regulator surfaces near the entangling points $a_{i}, b_{i}$ on the $z$ plane.  However, there is a subtlety regarding the mapping of the regulator surfaces.  If we choose the same regulator $\epsilon$ in the original $z$ plane for all entangling points, then the mapping to the double cover of $\mathcal{C}$ generically leads to cylinders of different lengths on each sheet (see figure \ref{cyl2} ).  We must account for this when computing the regulated length along the real $u$ direction.   To do this we lift to the double cover of $\mathcal{C}$ and compute the total length there of both sheets via
\bea\label{L}
L(\epsilon) = \sum_{j} \int_{(a_{j}+\epsilon, b_{j}-\epsilon)} \frac{du}{dz} dz  
\eea
Since this length already accounts for the two fermion fields, we have to divide the casimir energy by $1/2$.
By the same reasoning as in the single interval case, the thermal energy $E$ is again equal to the free energy. Adding these give the thermal/entanglement entropy
\bea\label{S}
S_{V}= \frac{1}{6}L(\epsilon)=\frac{1}{3} \left( \sum_{i,j} \log \frac{|a_{i}-b_{j}|}{\epsilon}-  \sum_{i<j}\log \frac{|a_{i}-a_{j}|}{\epsilon}-\sum_{i<j}  \log \frac{|b_{i}-b_{j}|}{\epsilon} \right)
\eea
 in agreement with \cite{casini2009reduced}.   Equations \eqref{L} and \eqref{S} generalizes to arbitrary number of intervals.    
 Finally, note that $ \beta (z)= \frac{2\pi }{u'(z)}$ defines a local entanglement temperature, from which we get the entanglement entropy density   
 \bea\label{SD}
 \frac{dS_{V}}{dz}=   \frac{c \pi}{3 \beta(z) } 
 \eea
 consistent with \eqref{S}.  This is completely analogous to the single interval case.

\section{Conclusion} 
In this work we have shown that the multi-interval modular flow can be obtained by coupling single-interval modular flow via an orbifold procedure and the associated twisted boundary conditions.  For a generic CFT reduced to $n$ intervals, the orbifold procedure is made possible due to a discrete, non-local $Z_{n}$ symmetry that mixes the $n$ intervals.    However, for free fermions, we find that this $Z_{n}$ symmetry is enhanced to a non-local $U(1)$ symmetry, which in turn leads to a non-local conformal symmetry for the reduced density matrix.   We defined a modular action $S_{V}$, and a corresponding finite temperature field theory in which this conformal symmetry is manifest.  The modular' stress tensor $T_{V}$ of this non-local CFT then determines the multi-interval modular Hamiltonian via the same formula  \ref{mod} as the single interval case.  Finally, we showed how this non-local conformal invariance is responsible for a particularly simple form for the entanglement entropy, which can be obtained by integrating a local entropy density \eqref{SD}.  

The natural question which emerges is whether this orbifold procedure can be usefully applied to a generic CFT.  As mentioned above, the $Z_{n}$ symmetry mixing the $n$ intervals is preserved in a general CFT, and is made manifest by the mapping \eqref{w}.   However,  the $U(1)$ non-local symmetry is generically broken, as one can see by adding a four fermion interaction to the free fermion \footnote{We thank Ben Freivogel for bringing this to our attention. }.  This means that the orbifold theory defined by the twisted boundary conditions is no-longer conformal.   Therefore, the conformal map we used to obtain the free fermion EE is not generically valid, which explains why the multi-interval EE generically differs from the free fermion value.  Nevertheless, it would be interesting to investigate deformations of the free fermion orbifold CFT and see whether these provide useful descriptions of the reduced density matrix for other CFT's.  In other words, one could map out  the RG flow in the space of non-local orbifold theories, for which the free fermion reduced density matrix is a fixed point.   

\begin{figure}[h]
\centering 
\includegraphics[scale=.35]{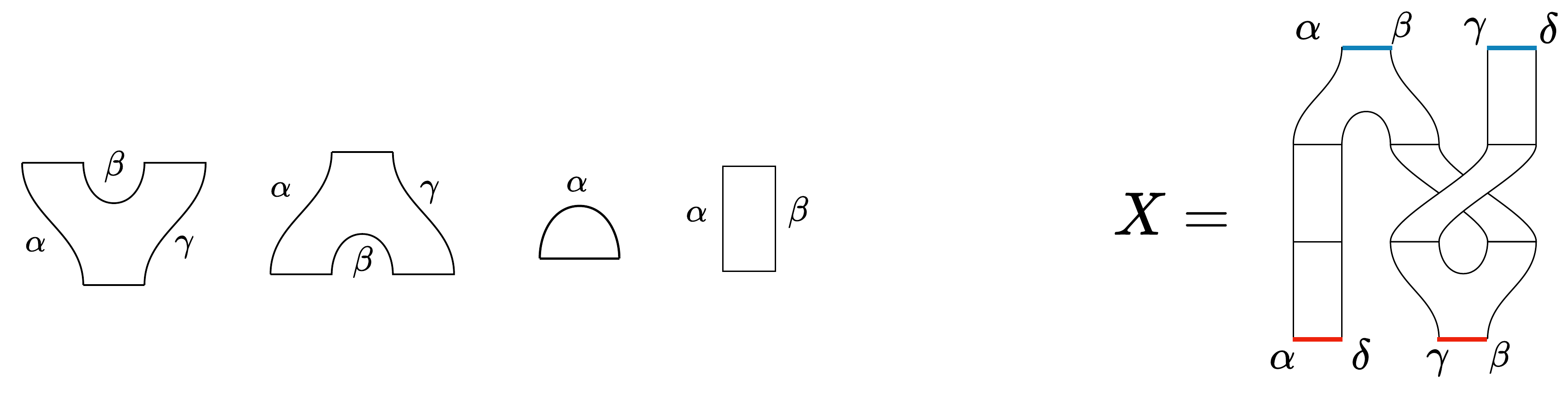}
\caption{On the left are the elementary cobordisms describing the modular evolution of intervals labelled by boundary conditions on the endpoints.   The cutting and regluing operator $X$ of \eqref{rhoV} can be expressed by composing the elementary cobordisms as shown on the right. }
\label{cobordisms}
\end{figure}

Finally, we wish to make a connection to recent work describing modular flows in terms of cobordisms of an extended quantum field theory.  A cobordism is a linear map that describes a quantum evolution between codimension 1 manifolds.   In the case of multi-interval modular flow, these codimension 1 manifolds represent the region $V$ at initial and final modular time, and the cobordism describes their splitting and joining.   A general cobordism between disjoint intervals is obtained by composing a set of elementary cobordisms (left figure in  \ref{cobordisms}) which includes  the basic ``open  string " splitting and joining interactions of a  BCFT: these correspond to the OPE's of boundary condition changing operators.   In terms of the elementary cobordisms, the cutting and regluing operator $X$ inserted at $z_{+}$  in the two interval modular flow is given by the cobordism in the right of figure \ref{cobordisms}, which describes a smooth evolution from the initial pair of intervals to their complement.  Thus, the cobordism interpretation  directly relates the two interval  modular flow to the basic data of the BCFT.    

\section{Acknowledgements}
It's a pleasure to thank Rob Myers, John Cardy, Diana Vaman, Israel Klich, Joan Simon, Will Donnelly, Janet Hung, Alice Bernamonti, Federico Galli, and Hal Haggard for many useful discussions.   This work was supported in part by the DOE grant DE-SC0007894, by Perimeter Institute for Theoretical Physics, Fudan University and the Thousand Young Talents Program.  Research at Perimeter Institute is supported by the Government of Canada through the Department of Innovation, Science and Economic Development Canada and by the Province of Ontario through the Ministry of Research, Innovation and Science.

\bibliographystyle{utphys}
\bibliography{RefsForEHpapers}

\providecommand{\href}[2]{#2}\begingroup\raggedright\begin{thebibliography}{10}

\bibitem{rehren2013multilocal}
K.-H. Rehren and G.~Tedesco, ``Multilocal fermionization,'' {\em Lett. Math.
  Phys.} {\bfseries 103} no.~1, (2013) 19--36.

\bibitem{Faulkner:2017vdd}
T.~Faulkner and A.~Lewkowycz, ``{Bulk locality from modular flow},''
  \href{http://dx.doi.org/10.1007/JHEP07(2017)151}{{\em JHEP} {\bfseries 07}
  (2017) 151},
\href{http://arxiv.org/abs/1704.05464}{{\ttfamily arXiv:1704.05464 [hep-th]}}.

\bibitem{Kabat:2017rw}
D.~Kabat and G.~Lifschytz, ``Local bulk physics from intersecting modular
  hamiltonians,'' \href{http://arxiv.org/abs/1703.06523}{{\ttfamily
  1703.06523}}. \url{https://arxiv.org/abs/1703.06523}.

\bibitem{Wong:2017pdm}
G.~Wong, ``{A note on entanglement edge modes in Chern Simons theory},''
\href{http://arxiv.org/abs/1706.04666}{{\ttfamily arXiv:1706.04666 [hep-th]}}.

\bibitem{Balakrishnan:2017bjg}
S.~Balakrishnan, T.~Faulkner, Z.~U. Khandker, and H.~Wang, ``{A General Proof
  of the Quantum Null Energy Condition},''
\href{http://arxiv.org/abs/1706.09432}{{\ttfamily arXiv:1706.09432 [hep-th]}}.

\bibitem{bisognano1975duality}
J.~J. Bisognano and E.~H. Wichmann, ``On the duality condition for a hermitian
  scalar field,'' {\em Journal of Mathematical Physics} {\bfseries 16} (1975)
  985.

\bibitem{myers}
H.~Casini, M.~Huerta, and R.~C. Myers, ``{Towards a derivation of holographic
  entanglement entropy},''
{\em JHEP} {\bfseries 1105} (2011) 036.

\bibitem{Casini:2017roe}
H.~Casini, E.~Teste, and G.~Torroba, ``{Modular Hamiltonians on the null plane
  and the Markov property of the vacuum state},''
  \href{http://dx.doi.org/10.1088/1751-8121/aa7eaa}{{\em J. Phys.} {\bfseries
  A50} no.~36, (2017) 364001},
\href{http://arxiv.org/abs/1703.10656}{{\ttfamily arXiv:1703.10656 [hep-th]}}.

\bibitem{casini2009reduced}
H.~Casini and M.~Huerta, ``Reduced density matrix and internal dynamics for
  multicomponent regions,'' {\em Class. Quant. Grav.} {\bfseries 26} no.~18,
  (2009) 185005.

\bibitem{Klich:2015ina}
I.~Klich, D.~Vaman, and G.~Wong, ``{Entanglement Hamiltonians for chiral
  fermions with zero modes},''
  \href{http://dx.doi.org/10.1103/PhysRevLett.119.120401}{{\em Phys. Rev.
  Lett.} {\bfseries 119} no.~12, (2017) 120401},
\href{http://arxiv.org/abs/1501.00482}{{\ttfamily arXiv:1501.00482
  [cond-mat.stat-mech]}}.

\bibitem{Cardy:2016yq}
J.~Cardy and E.~Tonni, ``Entanglement hamiltonians in two-dimensional conformal
  field theory,'' \href{http://arxiv.org/abs/1608.01283}{{\ttfamily
  1608.01283}}. \url{https://arxiv.org/abs/1608.01283}.

\bibitem{casini2005entanglement}
H.~Casini, C.~Fosco, and M.~Huerta, ``Entanglement and alpha entropies for a
  massive dirac field in two dimensions,'' {\em J. Stat. Mech.} {\bfseries
  2005} no.~07, (2005) P07007.

\bibitem{Wong:2013fk}
G.~Wong, I.~Klich, L.~Zayas, and D.~Vaman, ``Entanglement temperature and
  entanglement entropy of excited states,'' {\em JHEP} {\bfseries 2013} no.~12,
  (2013) 1--24.

\end{thebibliography}\endgroup

\end{document}